\documentclass[a4paper,11pt]{article}
\pdfoutput=1
\usepackage[english]{babel}
\usepackage{hyperref} 
\usepackage[pdftex]{graphicx}

\begin{document}

\pagenumbering{arabic}
\titlepage {
\title{Proposal to search for a dark photon in positron on target collisions at DA$\Phi$NE linac.}
\author{Mauro Raggi$^1$\footnote{mauro.raggi@lnf.infn.it}, Venelin Kozhuharov$^{1,2}$\footnote{venelin.kozhuharov@cern.ch}\\ 
{\it $^1$Laboratori Nazionali di Frascati - INFN, Frascati (Rome), Italy } \\
{\it $^2$University of Sofia ``St. Kl. Ohridski'', Sofia, Bulgaria}
}
\date{}
\maketitle

\abstract{Photon-like particles are predicted in many extensions of the Standard Model. They have 
interactions similar to the photon, are vector bosons, and can be produced together with photons. 
The present paper proposes a search for such particles in the $ e^+e^- \to U\gamma$ process in a 
positron-on-target experiment,
exploiting the positron beam of the DA$\Phi$NE linac at 
the Laboratori Nazionali di Frascati, INFN. 
In one year of running
a sensitivity in the relative 
interaction strength down to $\sim10^{-6}$ 
is achievable, in the mass region from 2.5 MeV$<M_U<$ 20 MeV. 
The proposed experimental setup and the analysis technique is discussed. }
} 

\makeatletter{}\section{Introduction}

The Standard Model of particle physics triumphed in 2012 with the discovery of the Higgs boson. 
However it is still far from consideration as the ultimate theory explaining all physical phenomena. 
The existence of Dark Matter is one of the examples of its failures and the search 
for a feasible explanation of that phenomenon is at present a major goal in particle physics. 

Despite attaining the highest energy ever reached at accelerators, 
LHC has not been able to provide  evidence for new degrees of freedom. 
An alternative approach are 
high statistics 
and high precision measurements which are sensitive to tiny effects that have escaped detection so far. 
Such effects could 
originate from the existence 
of a hidden sector of particles \cite{bib:kin-mixing}, 
interacting through a messenger with the visible ones. 
This scenario is appealing because it provides an explanation for the excess of positrons 
in cosmic rays observed by PAMELA in 2008 \cite{bib:pamela2008} and 
recently confirmed by FERMI \cite{bib:fermi-pos} and AMS \cite{bib:ams-pos}, 
namely, that they are from the annihilation of dark matter particles. 
The lack of excess of antiprotons \cite{bib:pamela-antip} suggests that the mass of the messenger 
should be below 1 GeV or that it interacts mainly with leptons. 
In addition, this messenger could 
provide the missing contribution to the present 
three sigma discrepancy between experiment and theory in the 
muon anomalous magnetic moment $a_{\mu} = (g_{\mu}-2)/2$ \cite{bib:g-2-discrepancy}. 

The simplest hidden sector model just introduces one extra U(1) gauge symmetry and a corresponding 
gauge boson: the ``dark photon'' or U boson. As in QED, this will generate interactions of the types
\begin{equation}
 \mathcal{L} ~\sim ~ g' q_f \bar{\psi}_f\gamma^{\mu}\psi_f U'_{\mu},
\label{eq:u1}
\end{equation}
where $g'$ is the universal coupling constant of the new interaction and 
$q_f$ are the corresponding charges of the interacting fermions.
Not all the Standard Model particles need to be charged under this new 
U(1) symmetry thus leading in general to a different (and sometimes vanishing) 
interaction strength for quarks and leptons. In the case of zero U(1) charge 
of the quarks \cite{bib:u1-gauge}, the new gauge boson cannot be directly 
produced in hadron collisions or meson decays. 

The coupling constant and the charges can result from a direct interaction between the 
Standard Model fermions and the new gauge fields or can be generated effectively 
through the so called kinetic mixing mechanism between the QED and the new 
U(1) gauge bosons \cite{bib:kin-mixing}. 
In the latter case the charges $q_f$ in eq.(\ref{eq:u1}) will be just 
proportional to the electric charge and the associated mixing term in the QED Lagrangian will be 
\begin{equation}
\mathcal{L}_{mix}=-\frac{\epsilon}{2}F^{QED}_{\mu\nu}F_{dark}^{\mu\nu}.
\end{equation}
The associated mixing coupling constant, $\epsilon$, can be so small ($<10^{-3}$) 
as to preclude the discovery of the dark photon in most of the experiments carried out so far. 
Another possibility is mass mixing with the Z, 
in which case the particle could also have Z-like properties. 

In the hypothesis that the dark sector does not contain any particle 
lighter than
the U boson, and that $M_U$ is smaller than twice the muon mass, 
the U can only decay to $e^+e^-$ pairs. 
In this case 
the U is expected to be a very narrow resonance with a total decay width  given by:
\begin{equation}
\Gamma_{U}=\Gamma_{U\to e+e-}=\frac{1}{3}\alpha \epsilon ^2 M_U \sqrt{1-\frac{4m_e^2}{M_U^2}} \left( 1+ \frac{2m_e^2}{M_U^2} \right)
\label{eq:u-width}
\end{equation}
which leads to a lifetime $\tau_U$ proportional to $1/(\epsilon^2M_U)$. 

\begin{figure}[t]
\centering
\includegraphics[width=13cm]{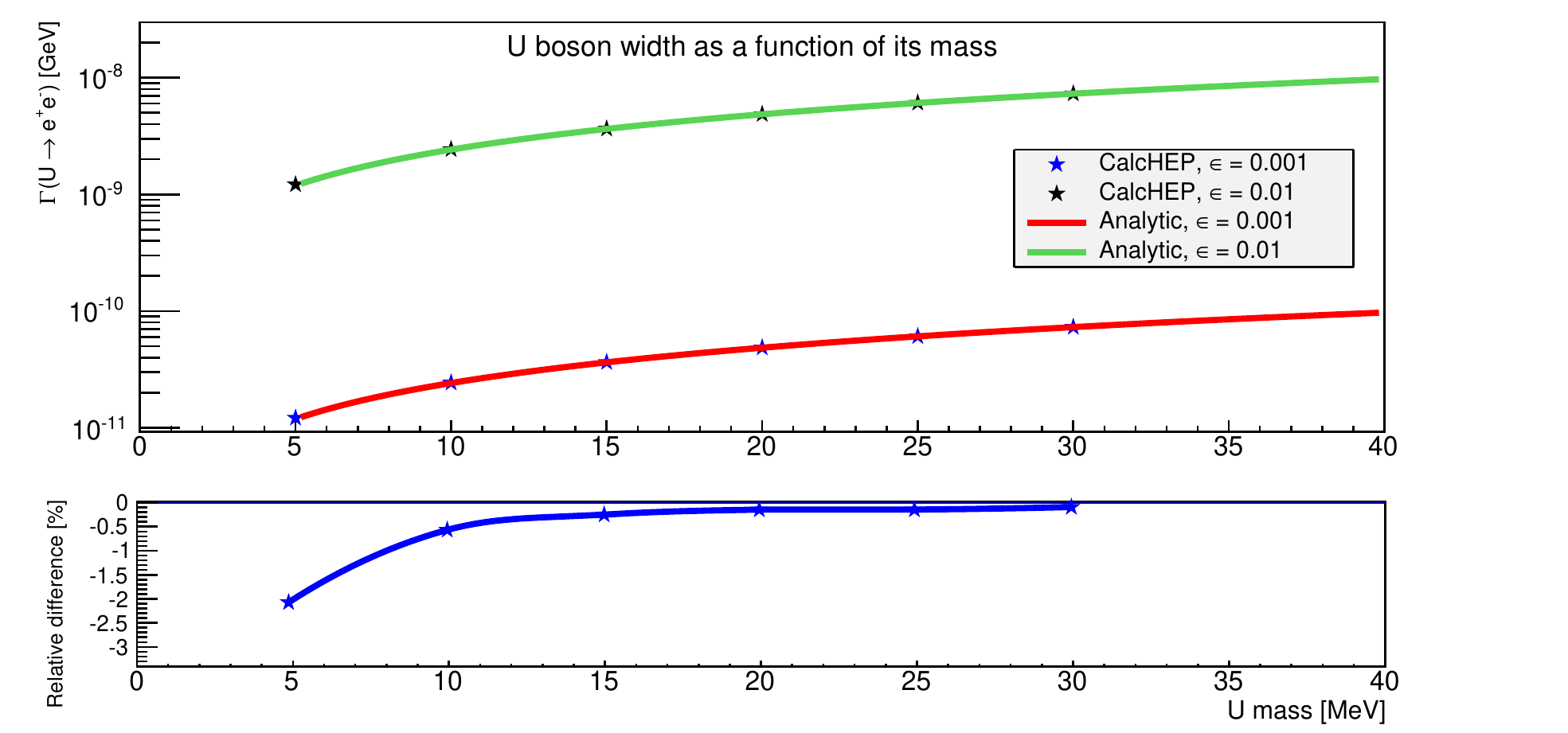}
\caption{U boson width calculated with formula \ref{eq:u-width} and compared to the toy model implemented in CalcHEP. 
Matching between the different calculation is better than 2.5\%.}
\label{fig:uboson-width}
\end{figure}

The present article describes a proposal devoted to the search 
for such a particle in a positron on target experiment. 
To be able to simulate the production of the U boson, the Lagrangian term from formula 
(\ref{eq:u1}), assuming $\psi = e$, was implemented in the CalcHEP \cite{bib:calchep} 
simulation software with $g' q_f = \epsilon$. The  results obtained for the decay width $\Gamma_{U\to e+e-}$ 
as a function of the U boson mass
were compared with the analytic formula (\ref{eq:u-width}) and are shown in Figure \ref{fig:uboson-width} 
for two different values of $\epsilon$. The difference between analytic and simulated values is less than 2.5\%,
validating the usage of CalcHEP for the calculation of the cross sections
 of other processes 
involving the U-boson.
In this notation the ratio of the  strengths
of the electromagnetic ($\alpha$) and the new U(1) ($\alpha'$) interactions is $\alpha'/\alpha = \epsilon^2$.

\makeatletter{}\section{Present status of experimental searches}
\label{sec:pres-stat}

Many searches have been performed in recent years to detect low energy exotic particles like the U boson 
exploiting different techniques that can be divided into three groups:
\begin{itemize}
\item Direct searches in beam dump experiments
\item Fixed target experiments
\item Direct searches in decays of mesons like $\pi^0$, $\eta$, $\phi$, $\Upsilon$
\end{itemize}
 
Based on the final state the experiment is looking at,
the searches can also be divided into visible and invisible. 
Visible searches implement a full reconstruction of the U boson decay products, 
usually $e^+ e^-$  or $\mu^+ \mu^-$ and are thus 
less demanding in terms of the definition of the initial-state kinematics.
The invisible searches do not assume that the U boson decay products, 
if any, are detectable by the experiment. 
However, if there are no Standard Model particles in the final state, 
a more stringent limit could be 
obtained through more efficient background vetoing. 
Both types of searches are complementary and are 
equally important.  

\begin{figure}[!htb]
  \begin{minipage}[t]{0.48\textwidth}
    \centering\includegraphics[width=\textwidth]{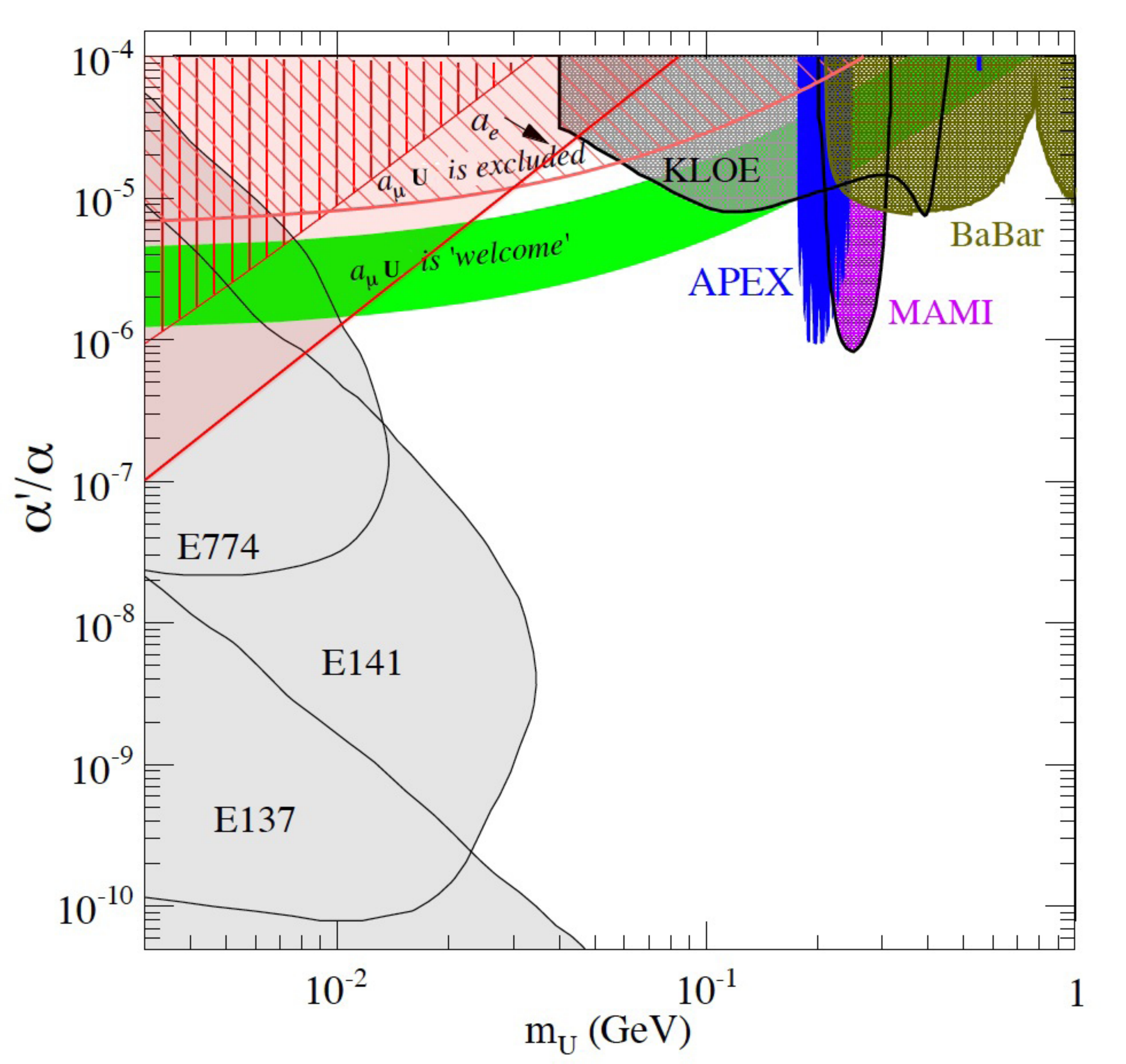}
    \caption{\it Exclusion region in the hypothesis of decay into lepton pairs}
    \label{fig:ExclVis}
  \end{minipage}\hfill
  \begin{minipage}[t]{0.48\textwidth}
    \centering\includegraphics[width=\textwidth]{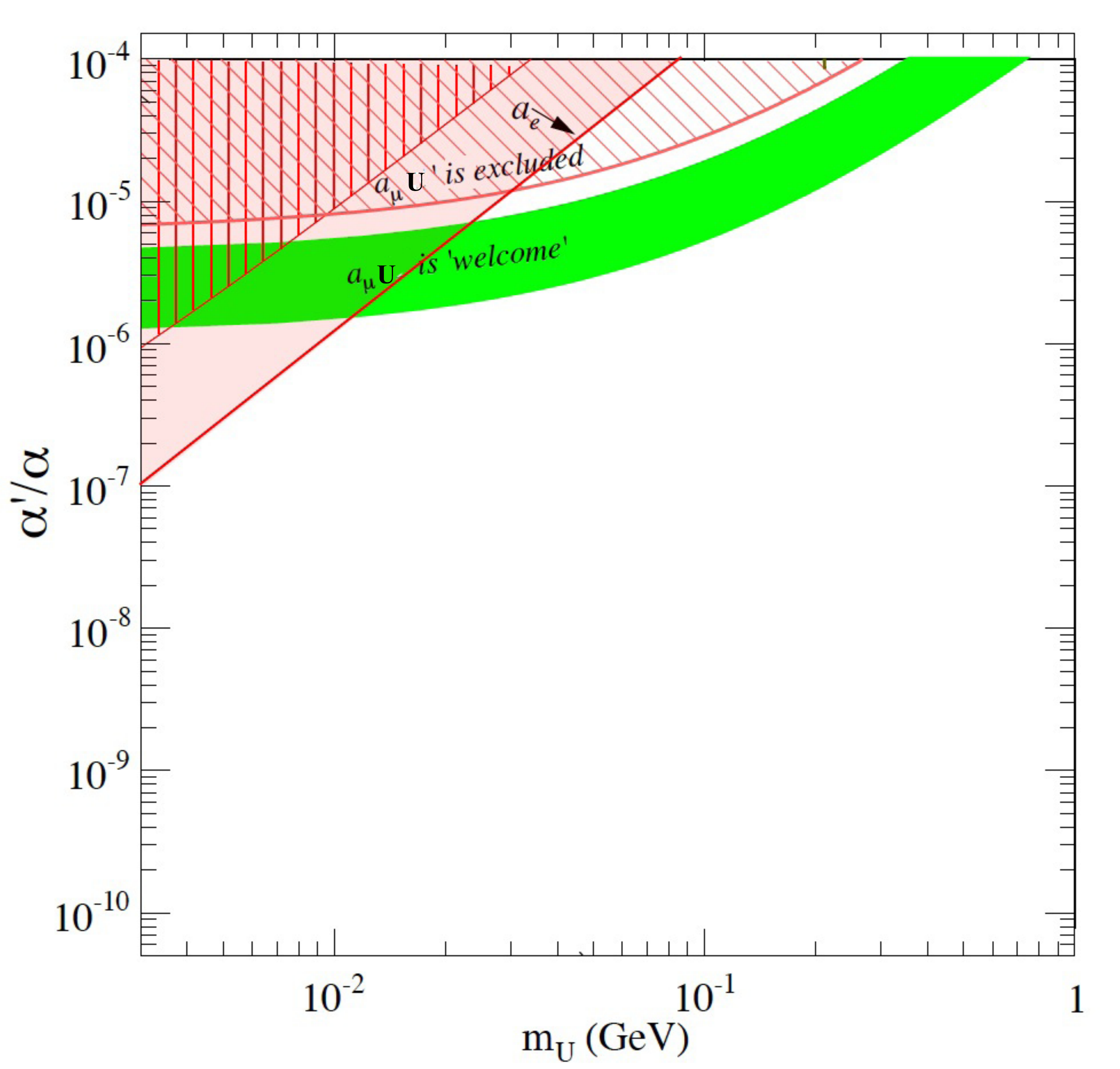}
    \caption{\it Exclusion region in the hypothesis of invisible U-boson, coupling to leptons}
    \label{fig:ExclInv}
  \end{minipage}
\end{figure}

The limits on U boson searches from beam dump experiments come from 
reanalysis \cite{bib:sarah-dark} of a compete set of experiments performed at SLAC and Fermilab during the '80s and '90s 
to search for light Axion-like particles \cite{DUMP1, bib:E137-dark, E774}.
These studies were triggered by the the observation of a 1.8 MeV 
mono-energetic positron  peak in heavy-ion collisions at GSI in 1983 \cite{bib:gsi-pos-peak}.
In these experiments, U bosons or axion like particles are produced, with a process similar to bremsstrahlung
of ordinary photons, by a very intense electron beam incident on a dump.  
The produced particles travel through the dump due to their long 
lifetime and are observed by a detector behind the dump through their decays into $e^+ e^-$ pairs. 
In this kind of experiment,
the measured quantity is in fact $\epsilon^2\cdot BR(U \to e^+e^-)$. 
If in the dark sector no particle lighter than the electron exists the limits are valid;
otherwise the U boson will decay into this new particle and may remain undetected in the experiment. 
In Figure \ref{fig:ExclVis} the present limits for dump experiments are shown in grey in the leftmost part. 

Fixed target experiments share the same beam type and production mechanism as dump experiments. 
The $e^+e^-$ invariant mass spectrum is searched for narrow resonances. 
The thickness of the target is reduced in order to 
allow short living U-boson to escape. The angular coverage of the spectrometer limits the acceptance and the 
accessible mass region. The exclusions from  MAMI\cite{oai:arXiv.org:1101.4091} and APEX\cite{APEX} experiments
are shown in Figure \ref{fig:ExclVis}.

Collider experiments have searched for resonances in the $e^+e^-$ ($\mu^+\mu^-$) mass 
spectra \cite{bib:bossi-dp, bib:essig-col} in the decay products of mesons
like $\pi^0$, $\eta$, $\phi$ or $\Upsilon$. Searches of this kind 
have been performed by WASA-at-COSY\cite{Adlarson:2013eza}, 
KLOE\cite{Archilli:2011zc,bib:kloe-phi-u} and BaBar\cite{oai:arXiv.org:0905.4539}. 
They are limited in mass reach by the mass of the meson, they assume that 
the U boson couples to quarks and to leptons with the
same strength, and it decays into lepton pairs. 
The results from collider experiments are shown in the top right part of Figure \ref{fig:ExclVis}.
 
The  U boson would also contribute to the anomalous magnetic moments of the electron and muon.
The contribution can be written as \cite{bib:ub-g2em}
\begin{equation}
 a = \frac{g - 2}{2}; ~ \Delta a = \frac{\alpha \epsilon^2}{2\pi} \times f,
\label{eq:g-2-limit}
\end{equation}
where $\alpha$ is the fine structure constant, 
$f = 1$ for $m_l \gg M_U$ and $f = 2m_l^2/(3M_U^2)$ for $m_l \ll M_U$. 
At present, $\alpha$ is in fact determined from the measurement of the electron 
magnetic moment $g_e$ \cite{bib:g-2-elec} and then is used to calculate 
the muon anomalous magnetic moment. Comparison between the theoretical  and the
experimental value allows a limit to be set in the $M_U - \epsilon^2$ parameter space, 
shown with diagonal stripes in Figure \ref{fig:ExclVis}. 
On the contrary the well known $3 \sigma$ discrepancy between the measured value of $g_{\mu}$ and 
the Standard Model prediction \cite{bib:g-2-discrepancy} 
can be 
explained by the existence of the U boson.
In this hypothesis a band of mass coupling values, in green in 
Figure \ref{fig:ExclVis}, is identified. This region is of particular 
interest and is not thoroughly explored at present.

An external knowledge of $\alpha$ is necessary to exploit the measurement of $g_e$ to set 
limits on the U boson parameters. The fine structure constant was extracted from 
a recent single measurement of the ratio between the Planck constant and the 
mass of the $^{87}$Rb atom \cite{bib:alpha-rb}. 
The precision achieved 
is an order of magnitude better than that for
 the previous measurements, which had been 
conducted two years earlier. 
That measurement, combined with equation (\ref{eq:g-2-limit}) and the latest tenth-order 
calculations of the QED contributions \cite{bib:ge-2-qed} 
to $g_e$,
gives a bound, shown by the pink line in Figure \ref{fig:ExclVis}, 
reaching $\epsilon^2 < 10^{-7}$ for $M_U = 1MeV$ and $\epsilon^2 < 10^{-4}$ for $M_U = 100MeV$. 
The limits from the measurements before \cite{bib:alpha-rb} are also shown in vertical stripes.

There are few studies on the searches of a U-boson not decaying into Standard
model particles. 
The data by SLAC Millicharge Experiment were interpreted within this scenario \cite{bib:mQ}.
The U-boson was assumed to decay into $\chi\chi$ and the $\chi$ to be detected 
by elastic scattering on atomic nuclei (through a virtual U-boson exchange).
The exclusions however depend on additional parameters, namely the dark matter mass and 
coupling $M_{\chi}$ and $~\alpha_{D}$ and is highly model dependent. 
The U-boson decaying into any invisible final state can be constraint 
also from the branching fraction of the decays $K^+ \to \pi^+\nu\bar{\nu}$
 \cite{bib:ub-g2em, bib:dark-review},  assuming non zero coupling to quarks. 
The present limits are in the range of $M_U \sim 100$ MeV 
and $\epsilon^2 \sim 10^{-6}$. 
Further improvements will be possible with the upcoming NA62 results \cite{bib:na62}.

The only limits on the $M_{U}$ and $\epsilon$ below 1 GeV that do not assume coupling
to quarks and are not dependent on the possible U-boson decay channel
are from the previously described
magnetic moment measurements. This scenario, being the most general one,
 is illustrated in Figure \ref{fig:ExclInv}.
Since the 
preferred  by $ (g_{\mu} - 2)$ region is not completely covered 
by direct searches even for visible decays, new experiments 
devoted to the the search for U boson are highly desirable. In addition, most of the searches 
of the invisible U-boson decays could be done with relatively 
small scale experiments \cite{bib:toro-new-dm-search}.

\makeatletter{}\section{Beam Test Facility at LNF}

\begin{figure}[ht]
\centering
\includegraphics[width=9cm,height=12cm,angle=270]{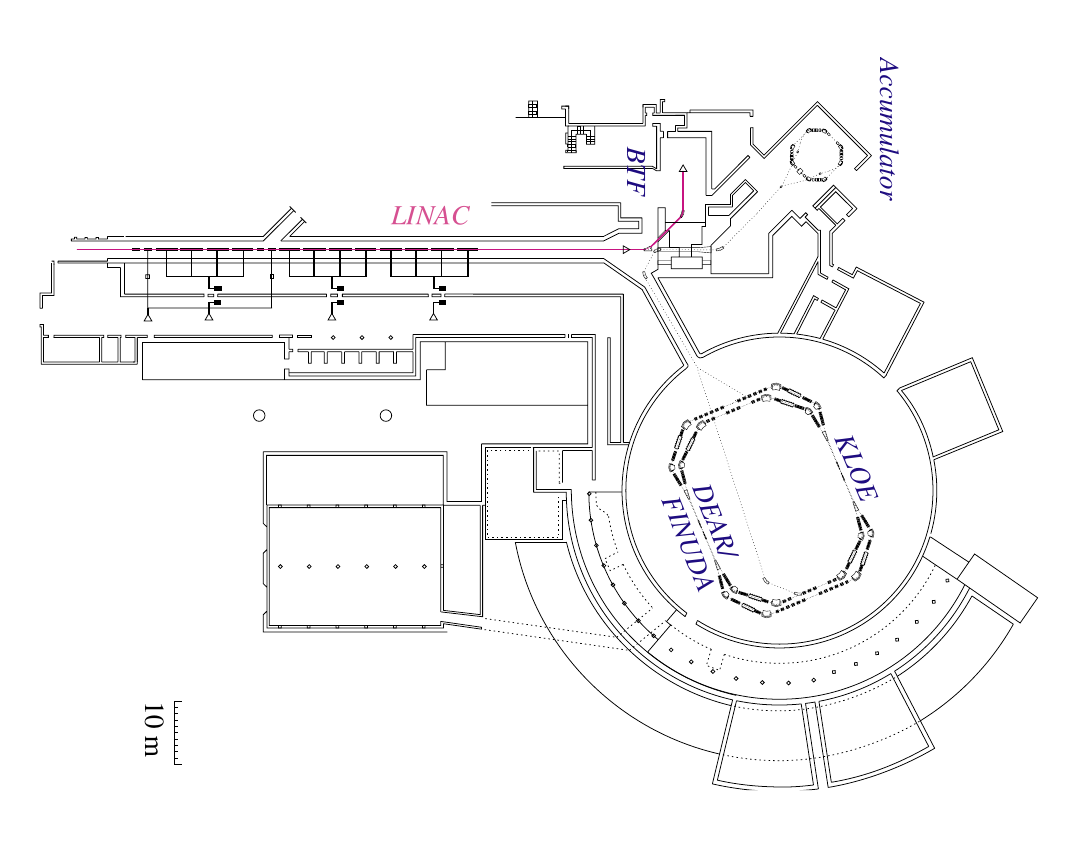}
\caption{Layout of the DA$\Phi$NE complex at Laboratori Nazionali di Frascati, INFN}
\label{fig:btf}
\end{figure}

The DA$\Phi$NE beam-test facility (BTF), shown in Figure \ref{fig:btf}, 
is a beam transfer-line \cite{Ghigo:2003gy} 
from the DA$\Phi$NE linac to an independent experimental hall with a dedicated control room. 
It is capable of providing  up to 50 bunches per second of electrons or 
positrons
with 800/550 MeV maximum energy.
The minimum energy 
is below 250 MeV both 
in electron and positron mode, even though some optimization is necessary 
in order to obtain stable 
operation.
An energy upgrade of the linac is foreseen within the next three years \cite{bib:btf-upg}, 
bringing the maximum energy for electrons/positrons to about 1050/800 MeV. 
Each bunch consists of microbunches with total length of 350 ps with 140 ps flat top. 
Beam granularity together with the subdetectors time resolution 
can be used to reduce the pile up background. 
The typical emittance of the electron/positron beam is of $1 (1.5)$ mm*mrad, 
while the maximum beam current is of 500 mA (100 mA). 
Beam characteristics (spot size, divergence, momentum resolution) 
strongly depend on the beam energy and the number of particles per bunch. 
The parameters are summarized in Table \ref{tab:btf}. 

Any of the 50 linac bunches can be injected either to the Accumulator, 
and from there to the DA$\Phi$NE main rings, or extracted (by a pulsed magnet) 
to the BTF beam line.
The BTF beam
can be attenuated by a variable-depth copper target and energy selected with an accuracy of better than 1\%
before being transported to the experimental hall. 
In the experimental hall the particles can be either delivered on a straight beam line, 
or deflected by 45 degrees by a dipole magnet. 
In normal operation, one of the 50 pulses/second is sent to a magnetic spectrometer 
for a precise measurement of the beam momentum.
\begin{table}
\begin{center}
  \begin{tabular}{|l|c|c|c|c|}
   \hline
	                     & $e^+$ & $e^-$ & $e^+_{Upg}$ & $e^-_{Upg}$\\
   \hline
   \hline
Maximal beam energy [MeV]    & 550  & 800  & 800  &  1050  \\
Beam rate [particles/bunch]  & $6.2\cdot10^8$  & $6.2\cdot10^8$  &  $6\cdot10^{9}$ & $3.1\cdot10^{10}$\\
Number of bunches per second [Hz] & 50   & 50  &  50 & 50 \\
Max. averaged current during a bunch [mA] & 10  & 10  & 100 & 500 \\
Typical emittance  [mm*mrad] &  1.5& 1.&1.5& 1.\\
Beam spot size ($\sigma$) [mm] &  2. & 2. & 2. & 2.\\
Bunch length [ns] &  10 & 10 & 40 & 40\\
\hline
\end{tabular}
\end{center}
\caption{BTF beam parameters \cite{Ghigo:2003gy}. A possible increase of the performance after an upgrade is also shown.}
\label{tab:btf}
\end{table}
Presently, the maximum beam intensity in the BTF experimental hall is limited 
by radio-protection safety regulations at $ 3.12\cdot10^{10}$ particles/second.

The most important limitation of the linac for the project described in this paper was the very 
short duration of the bunch, 10 ns, and the 50 bunches/s delivered. 
In fact, to obtain a sufficiently high event rate 
a high number of $e^+$ per bunch has 
to be used, increasing the pile up probability. This pile up could be partially
resolved by the subdetectors with a time resolution of the order of a nanosecond,
but this required an increased bunch length. 
A recent upgrade of the pulser of the linac gun \cite{bib:btf-gun-upg} now allows variation of 
the pulse duration between 2 and 40 ns, but there is an indication that the bunch 
width can be extended up to about 200 ns. The bunch rate, however, can not be increased at present.

The present proposal assumes that the BTF will be able to provide 
50 bunches per second with duration of 40 ns and $10^4$ - $10^5$ positrons.

\makeatletter{}\section{Experimental technique}

The experiment aims to search for the production of a U boson in the process 
\begin{equation}
 e^+e^- \to U \gamma,
\end{equation}
where the positrons are the beam particles and $e^-$ are the electrons in the target.
The accompanying photon is then detected by a calorimeter regardless of the U decay products. 
A single kinematic variable characterizing the process, the missing mass, is computed using the formula:

\begin{equation}
M_{miss}^2 = (P_{e^-}+P_{beam}-P_{\gamma})^2. 
\label{eq:mmiss}
\end{equation}

Its distribution should peak at $M_U^2$ for U boson decays, at zero for the concurrent
$e^+e^- \to \gamma\gamma$ process, and should be smooth for the remaining background. 
The approach described  provides sensitivity for both  visible and invisible searches.

The proposed experimental setup uses the BTF positron beam impinging on a thin target
and 
  is composed of the following parts:

\begin{itemize}
\item {\bf Active target}, to measure the average position of the beam during a single BTF spill
\item {\bf Spectrometer}, to measure the charged interaction products in a given momentum range
\item {\bf Dipole magnet}, to deflect the primary positron beam out of the spectrometer and calorimeter 
and to allow momentum analysis.
\item {\bf Vacuum chamber}, to minimize the unwanted interactions of primary and secondary particles.
\item {\bf Electromagnetic calorimeter}, to measure and/or veto final state photons.
\end{itemize}

\begin{figure}[htb]
\centering
\includegraphics[width=11cm]{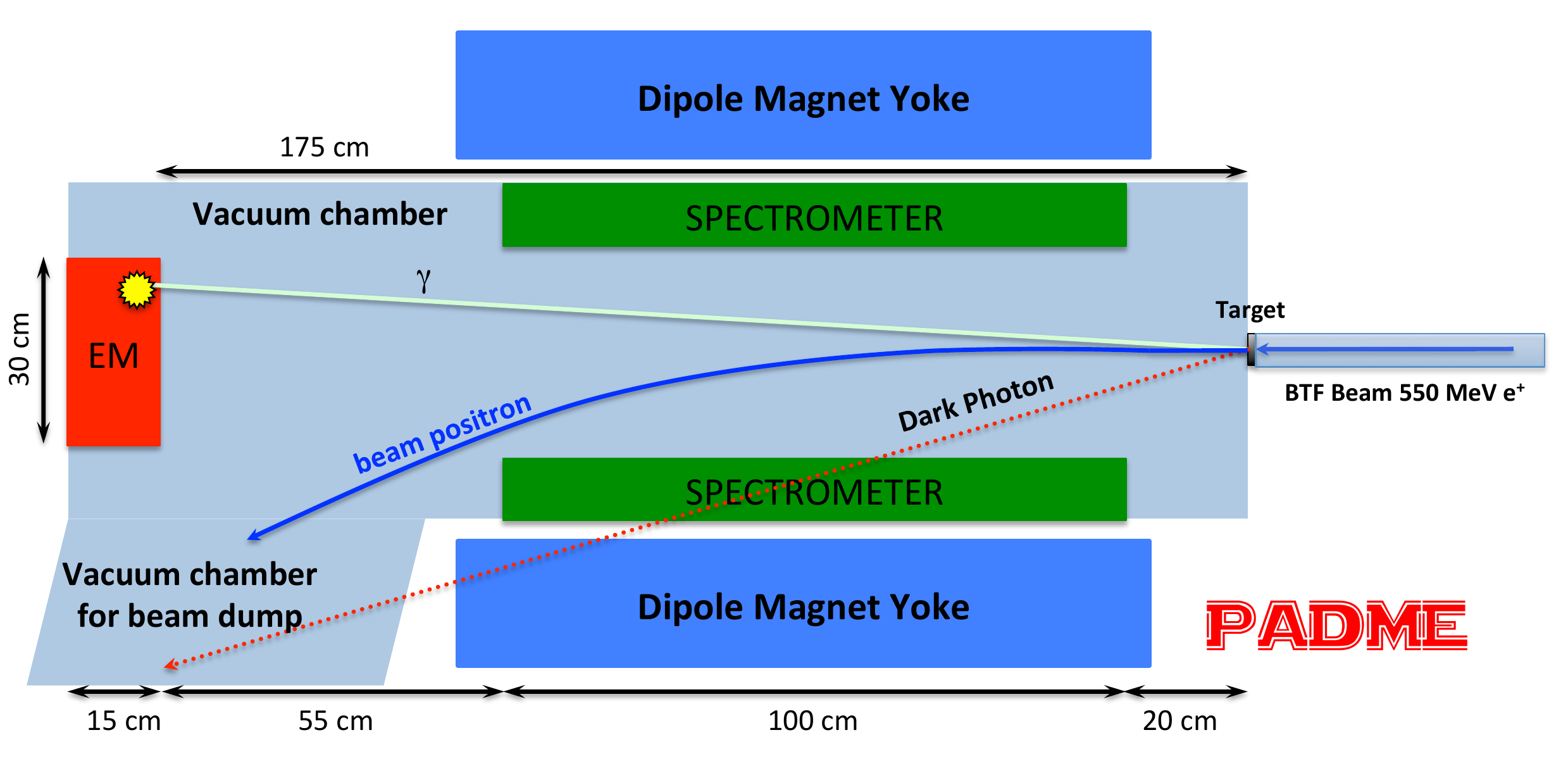}
\caption{Schematic of the Positron Annihilation into Dark Matter Experiment (PADME).}
\label{fig:Exp}
\end{figure}

A schematic view of the experimental apparatus is shown in Figure \ref{fig:Exp}.

The primary beam crosses the target and if a beam particle 
does not interact it is bent by the magnet in between the end of 
the spectrometer and the calorimeter, thus leaving the experiment undetected.
If any kind of interaction causes the positron to lose more than 50 MeV of energy the 
magnet bends it into the spectrometer acceptance, providing the veto 
capabilities against background. 
If a U boson decays into $e^+e^-$, the tracks are also detected by the spectrometer. 
This could be used to perform visible decay searches.

Due to the very thin target
the majority of beam does not interact. It is transported in 
vacuum to the end of the experimental setup 
 and can be reused for detector 
testing, provided that appropriate beam transport system is built. 

The proposed experiment is compatible with the operation of the DA$\Phi$NE ring. 
However due to the 40\% reduction of the available beam time due to injection into
the DA$\Phi$NE machine a longer data taking period or a reduction of the sensitivity should be foreseen. 
A standalone operation has the advantage of profiting from beam energy 
variation and nominal running conditions.

\subsection{U Boson production at BTF}

The possible U boson production mechanisms accessible in 
$e^+$-on-target collisions are $e^+e^- \to U \gamma$ and $e^+A \to e^+ A U $, 
the so called annihilation and U-strahlung production:

\begin{figure}[ht]
\centering
\includegraphics[width=11cm]{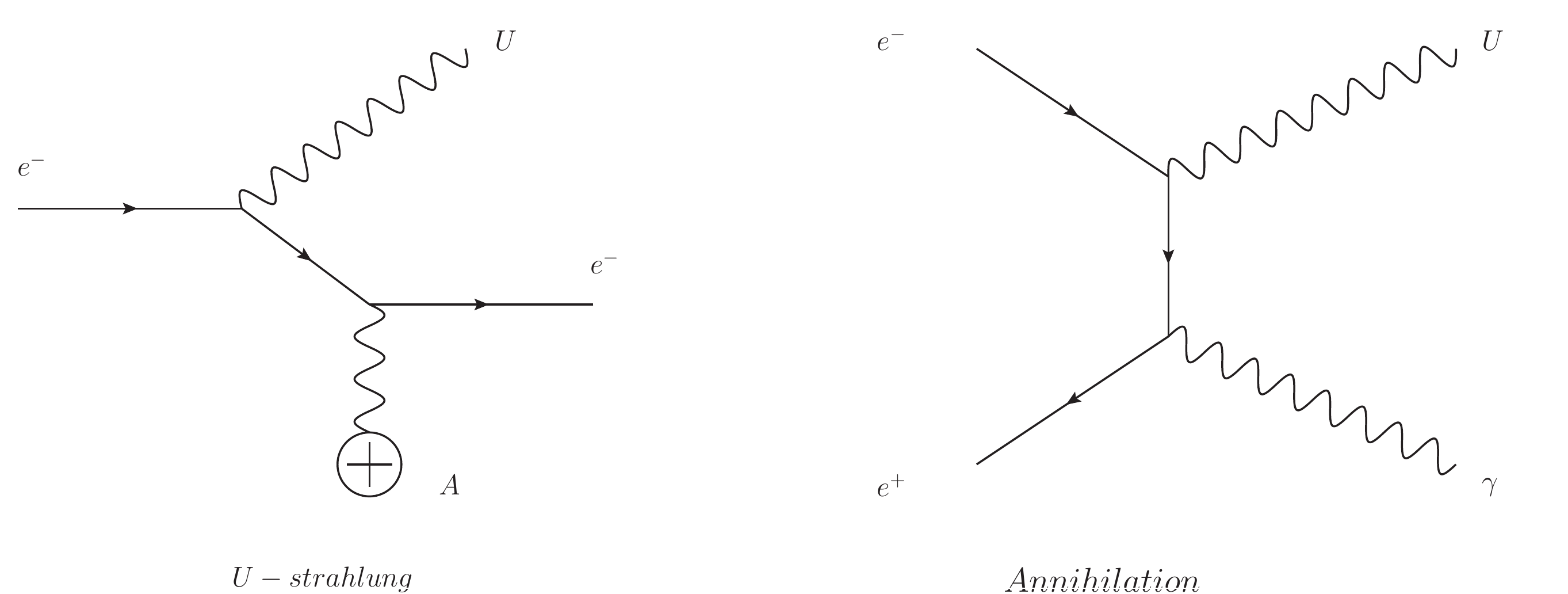}
\caption{U boson production mechanisms.}
\label{fig:Uprod}
\end{figure} 

Both process are similar to the ones for ordinary photons, 
as shown in Figure \ref{fig:Uprod}, and their cross section scales with $\epsilon^2$.
The present linac maximum positron energy of 550 MeV allows 
the production of U bosons through annihilation up to a centre of mass energy of 23.7 MeV. 
The kinematical constraints of the initial state is of great importance for rejecting the background. 

\begin{figure}[!htb]
  \begin{minipage}[t]{0.46\textwidth}
    \centering\includegraphics[width=\textwidth]{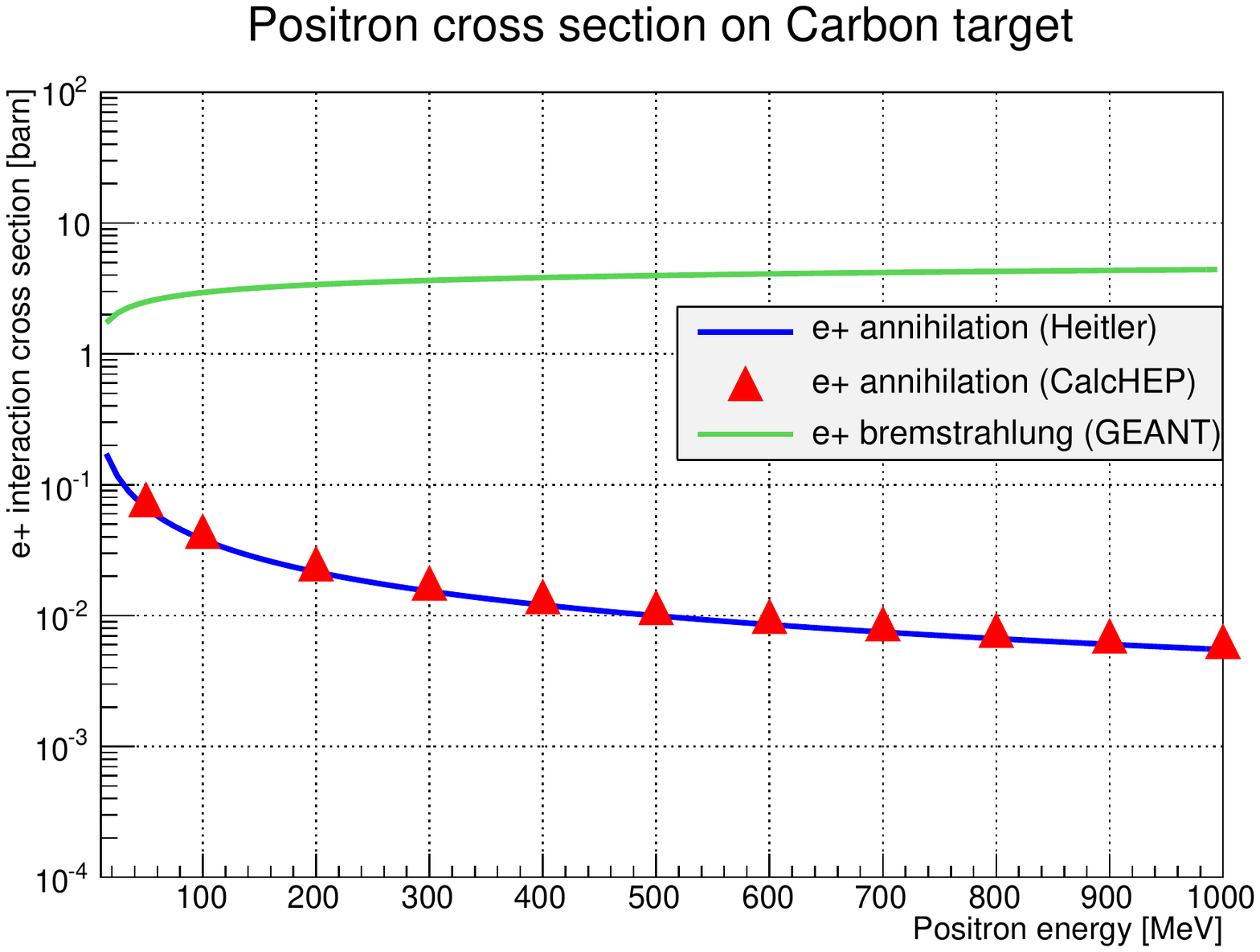}
    \caption{\it Positron cross section on a carbon target. }
    \label{fig:cross-section}
  \end{minipage}\hfill
  \begin{minipage}[t]{0.5\textwidth}
    \centering\includegraphics[width=\textwidth]{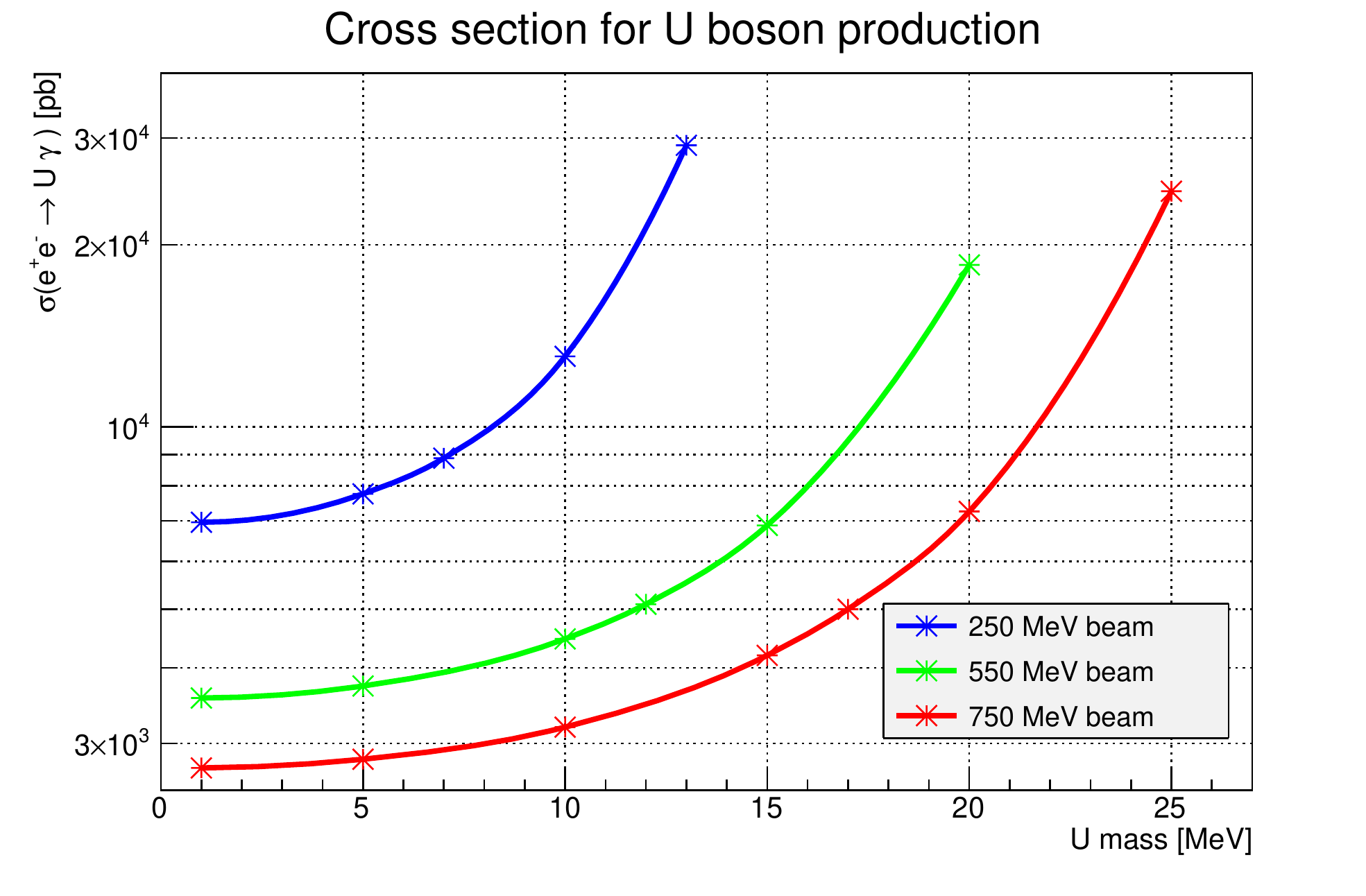}
    \caption{\it U boson production cross section (for $\epsilon = 10^{-3}$) as a function of mass for different beam energies }
    \label{fig:ee-ug-sigma}
  \end{minipage}
\end{figure}

The cross sections for annihilation and bremsstrahlung emission of an
ordinary photon with energy above 1 MeV for positrons on a carbon target 
are shown in Figure \ref{fig:cross-section}. The annihilation cross section was calculated directly with 
CalcHEP and was compared with the  Heitler formula  implemented in 
GEANT4\cite{bib:annih}. Agreement within 2\% is observed. 
For the bremsstrahlung the GEANT4 parametrization of the cross section was used. 

Since the ratio of annihilation to bremsstrahlung cross sections 
is proportional to $1/Z$, the 
lower the Z of the target the better. 
However due to the requirement to form a rigid and self supporting structure, 
carbon was chosen as a material, providing annihilation/bremsstrahlung ratio of $2.3*10^{-3}$  
($5.1*10^{-3}$)  for 550 MeV (250 MeV) positrons.
The cross section for annihilation of positrons with energies of 550 MeV (250 MeV) 
is $1.5~mb$ ($3~mb$  ) per free electron leading to a probability of 
$6*10^{-6}$ ($1.2*10^{-5}$) in a 50 $\mu m$ thick carbon target. 

Operating the beam line in the regime of $10^4-10^5$ $e^+$ per bunch and 
50 bunch/s in one year of data taking with 60\% efficiency, a sample of  
$\sim 10^{13} - 10^{14}$ positrons on target can be collected, 
corresponding to $\sim (6\cdot10^7 - 6 \cdot10^8)$ annihilation per year. 
In a zero background experiment, 
a limit down to  $10^{-8}-10^{-9}$ in $\epsilon^2$ could be set.

\begin{figure}[!htb]
  \begin{minipage}[t]{0.48\textwidth}
    \centering\includegraphics[width=\textwidth]{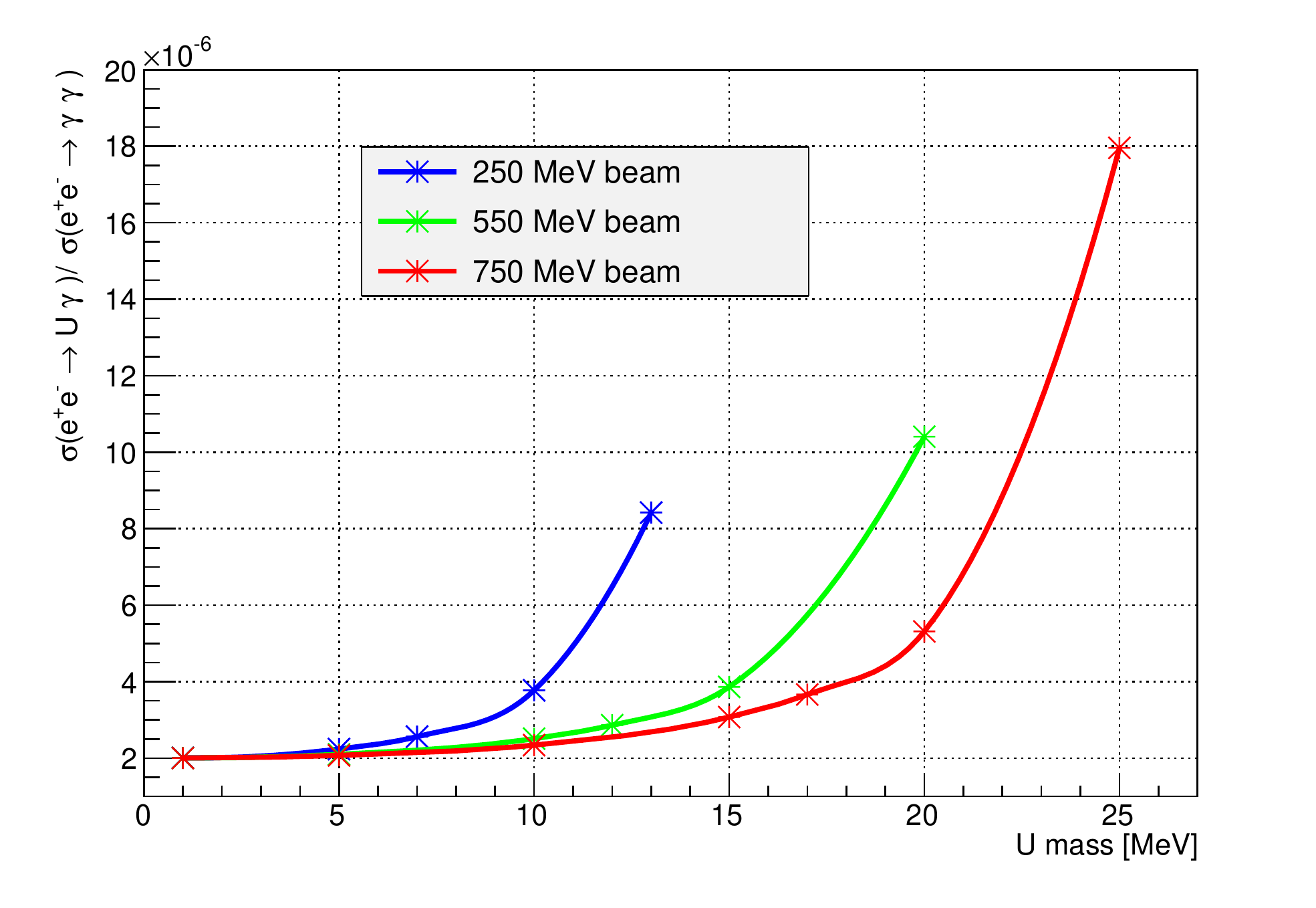}
    \caption{\it Ratio of the U boson production (for $\epsilon = 10^{-3}$) and two photon annihilation 
cross section as a function of the U boson mass for different beam energies}
    \label{fig:ee-ug-sigma-ratio}
  \end{minipage}\hfill
  \begin{minipage}[t]{0.48\textwidth}
    \centering\includegraphics[width=\textwidth]{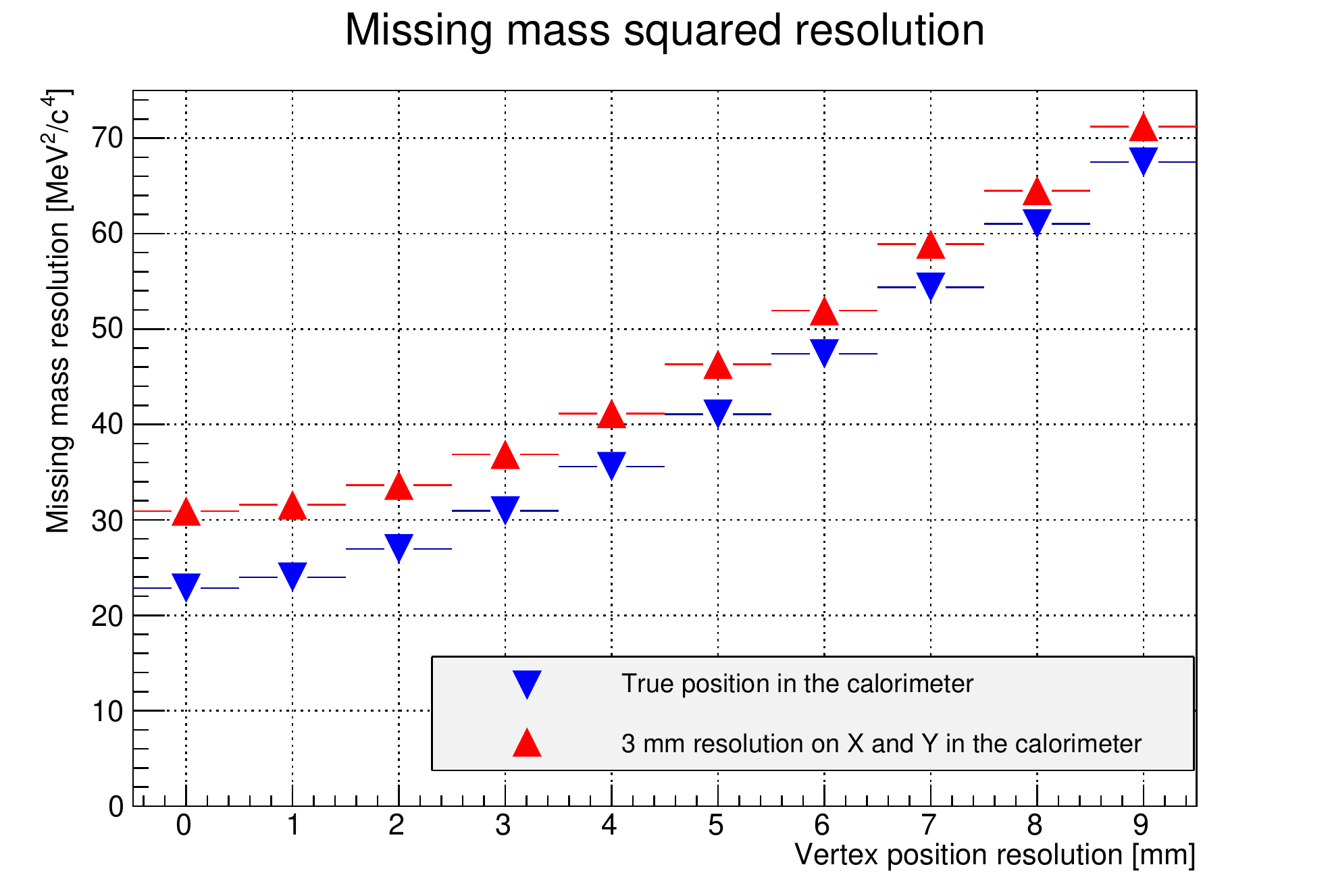}
    \caption{\it Dependence of the missing mass squared resolution on the vertex position resolution for a U-boson with $M_U= 15 MeV$.}
    \label{fig:mmiss-res-vtx}

  \end{minipage}
\end{figure}

Another advantage of the U-annihilation production process comes from the 
resonant enhancement of the U boson production cross section.
The production cross section, shown for different beam energies in Figure \ref{fig:ee-ug-sigma},
and the $\sigma(U\gamma)/\sigma(\gamma\gamma)$, shown in Figure \ref{fig:ee-ug-sigma-ratio} 
rise fast when the mass of the 
U boson approaches the available CM energy. 
The ratio of the the annihilation to bremsstrahlung cross section also increases
with the decrease of beam energy, (as seen in Figure \ref{fig:cross-section}) giving an additional rise to the production
of the U-boson.
This can  be exploited by an experiment at the the BTF beam line due to its variable 
beam energy and could enhance the sensitivity to low-mass U-bosons.

\subsection{Active target}

Scattering inside the target material worsens the knowledge 
of the momentum and  direction of the primary beam 
and broadens the reconstructed missing mass spectrum. 
In fact, in the invisible searches the kinematics cannot be closed 
without assumptions on the decay vertex and beam direction, while in the 
visible searches, the vertex position helps in rejecting fake tracks.
For this reason the usage of 50$\mu m$ target 
is proposed, for which the simulation showed that the relation 
$E_{\gamma, brems}+E_{e+}=E_{Beam}$ is fulfilled with a resolution better than 
the initial beam spread (1\%).
In addition, the probability of a single annihilation 
is $\sim$5\% per bunch with $\sim 10^4$ positrons.

\begin{figure}[!htb]
  \begin{minipage}[t]{0.48\textwidth}
\centering\includegraphics[width=6cm]{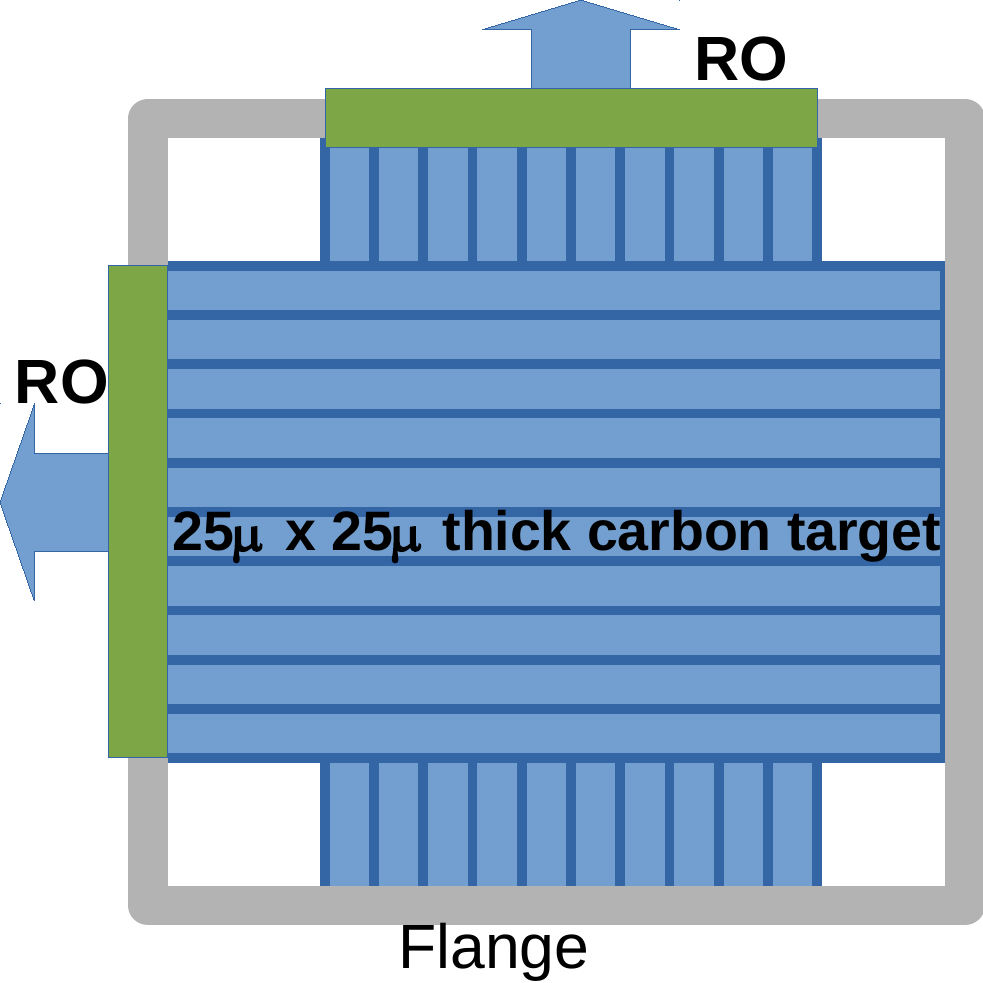}
\caption{Schematics of the beam monitor. A design with 2 mm strips both in horizontal and vertical direction are foreseen}
\label{fig:Monitor}

  \end{minipage}\hfill
  \begin{minipage}[t]{0.48\textwidth}
    \centering

\includegraphics[width=7cm]{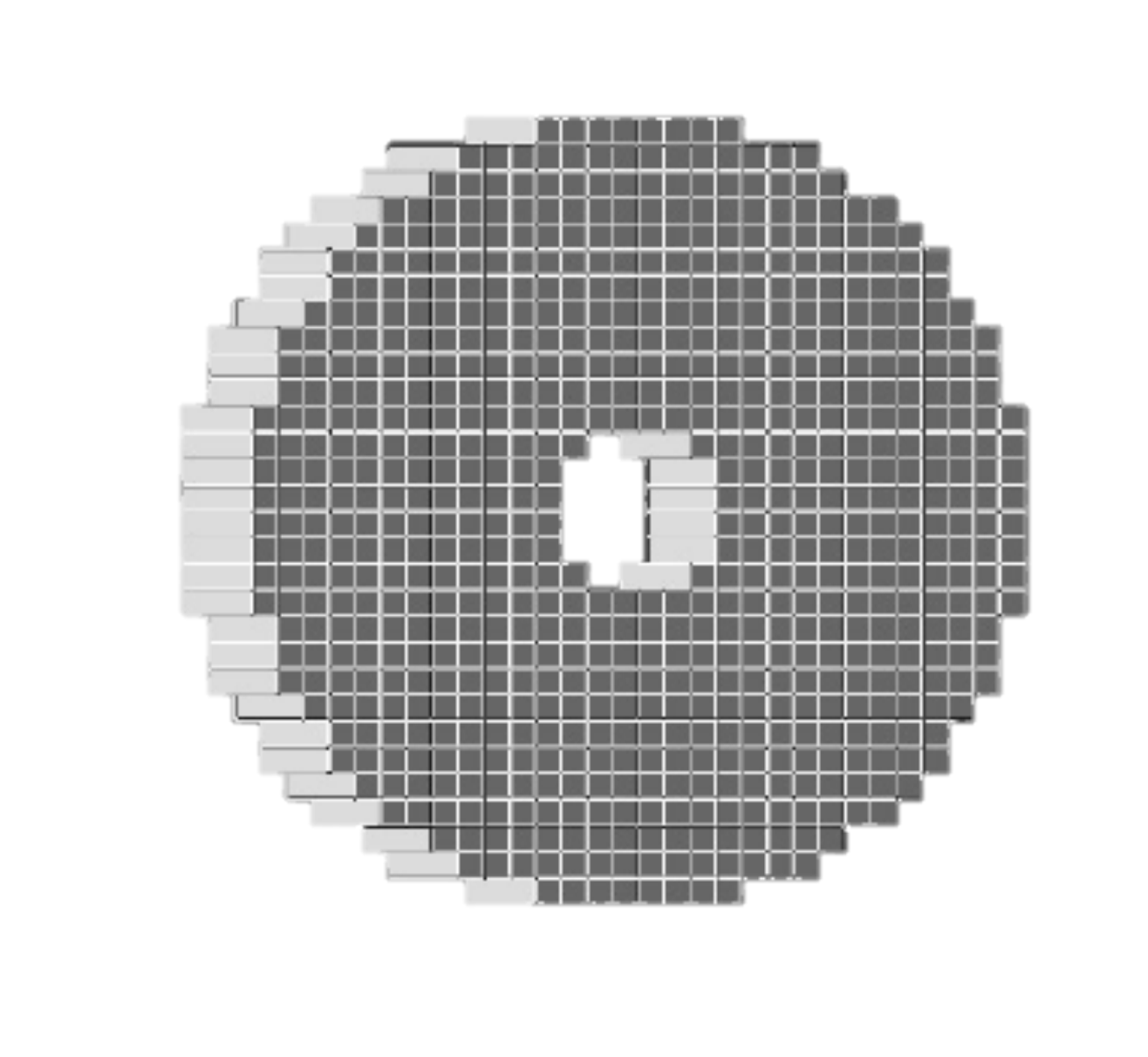}
\caption{GEANT4 model of the EM calorimeter.}
\label{fig:Calo}

  \end{minipage}
\end{figure}

The importance of the beam position measurement can be seen in Figure \ref{fig:mmiss-res-vtx}, 
where the missing mass resolution for a U boson with mass 15 MeV,
assuming perfect calorimeter positioning and a realistic 3mm 
resolution on the cluster position inside the calorimeter, is shown. 
In order to obtain missing mass resolution down to $\sim 30 MeV^2/c^4$ 
it is necessary to provide better than 
2mm position resolution on the beam spot on the target, which is used 
for the calculation of the angle of the emitted photon.
A single BTF bunch features such a small spot, but due to 
hysteresis and stability of the currents in the magnets, it is 
difficult to achieve long term positioning stability 
better than 5 mm \cite{Eigen:2013wka}.  

The design of the proposed target, shown schematically in Figure \ref{fig:Monitor},
consist of two orthogonal layers of ten strips made of carbon, 
each with dimensions
 2 mm $\times$ 50 mm $\times$  25 $\mu$m, mounted on a vacuum flange.
Both diamond and graphite could possibly be used due to their 
tolerance to radiation damage and good heat conductivity.
The latter is necessary to avoid having to use a cooling system 
while operating in high intensity beams. 

The target will also act  as a beam monitor detector, 
 providing the beam spot position bunch per bunch. 
Its readout will be based on measurement of the 
 secondary electron emission (graphite) or Cherenkov light (diamond). 
By reconstruction of the beam spot, the uncertainty on the beam geometry
will be reduced to that of only a single bunch. 
An online, two-dimensional beam image would also be useful 
to optimize the beam parameters while setting up the beam. 
In addition, it can also be used to measure 
the number of particles per bunch for intensities of $<10^5$ positron/bunch, 
at which methods based on induced current start to be ineffective.

\subsection{The dipole magnet}

After crossing the active target the charged remnants of the primary 
beam will be swept away by a sweeping magnet. 
Because the target is very thin, most of the beam particles will remain in the beam, 
in the worst case with slightly degraded energy. 
Due to the low energy of the beam and relaxed spatial constraints, the magnet can be a conventional one.
 Geant4 simulations show that a magnetic field of $\sim$0.6 Tesla 
is enough to deflect the primary 550 MeV beam out of the calorimeter 
acceptance while keeping most of the electrons from U boson decays inside of the spectrometer acceptance.
The magnet surrounds the vacuum region and contains the 
whole tracking detector, thus serving as a spectrometer magnet. 
A gap of $50$ cm between the coils and a uniform field of length $1$ m is assumed.

\subsection{The spectrometer}
The spectrometer role is to suppress by a factor of 100 the background from bremsstrahlung
events by detecting the primary beam positrons. 
A tracking detector inside the magnetic field
will be used to detect
charged particles and to  measure their momentum.  
The spectrometer could be either a cylindrical chamber with
20 cm inner radius and 25 cm outer radius or composed by  
two planar trackers, one on each of the positron and the electron deflection sides.
It is expected to measure the coordinates of a crossing track 
 with a precision of 300 $\mu m$ in each layer. 
The size of the spectrometer 
is defined by the condition that it should detect charged particles 
with momentum from 
50 MeV to 500 MeV  travelling along the beam axis with an efficiency of 99\%. 
Positron with momentum $\vec{p} = 550$ MeV  should be deflected outside its acceptance.
In order to avoid conversions of the emitted bremsstrahlung photons the tracker 
should be at least 20 cm far from the beam line.

The proposed detector is 
with five active layers and a total length of 1m. 
Few technologies are considered at present. A possible solution is a 
 triple GEM chamber if a thin vacuum pipe on the positron path 
doesn't worsen significantly the background rejection. An alternative option is a 
scintillating fibre tracker that could be operated in vacuum. A hybrid solution
(plastic scintillator for good time resolution and a GEM tracker) is also considered.

\subsection{The decay chamber}
Due to the high intensity of the beam
and the extremely thin target 
(0.04\% $X_0$), 
$e^+$ interactions in air can produce a significant contribution to the background. 
In fact, since the radiation length of air is 285 m at a pressure of 1013 mbar and 
the distance from the calorimeter to the target is $\sim$ 2 m, 
the atmospheric air thickness is $0.7\% X_0$, 
which is much larger than the thickness of the
target itself.
   
An MC simulation performed with $1$ mbar pressure showed 
a significant increase in the background 
with respect to an experiment in vacuum. 
In interactions in the residual air, the kinematical constraints 
(i.e. from missing mass) are weakened since the information on the interaction 
position and the initial beam momentum is completely lost. 
In addition, the emitted 
bremsstrahlung photons do not travel through the central calorimeter hole.
For the same reason, beam particles not interacting in the target 
must be transported in vacuum to 
the dump.
Due to the fact that the spectrometer itself contains gas  and will not intercept the primary beam, 
it can be placed outside the vacuum region. 
A vacuum with pressure $10^{-1}$ mbar was chosen for the experiment,
which makes the background contribution
from beam-air interactions negligible. 

\subsection{The electromagnetic calorimeter}

The electromagnetic calorimeter is responsible for the reconstruction of the photon 4-momentum. 
An energy resolution of better than $\sim5\%$ for photons with energies down to 100 MeV 
and a cluster position resolution of $3$ mm are needed in order to achieve 
squared missing mass resolution of 30 MeV$^2$/c$^4$. 
In addition, a compact shower development 
is desired to minimize overlapping
of signal and background photon clusters.

To achieve such performance, 
a highly-segmented, inorganic crystal calorimeter made of 
LYSO 
was chosen due to its characteristics, listed in Table \ref{tab:LYSO}. 
LYSO has high density, very short $X_0$ and small Moliere radius together 
with high light output and short decay time.
The calorimeter is an approximate cylinder with a diameter of 30 cm and depth 
of 15 cm filled with 1x1x15 cm$^3$ crystals with a round shaped 
central hole of 4 cm radius, as shown in Figure \ref{fig:Calo}. 
The active volume will be 9840 cm$^3$  for a total of 656 crystals.
Resolutions down to
\begin{equation}
\label{ eqn:CaloRes}
\sigma_E / E=\frac{1.1\%}{\sqrt{E}} \oplus \frac{0.4\%}{E} \oplus 1.2\%
\end{equation}
have been recently obtained in the R\&D for the calorimeter of the SuperB project in tests with LYSO crystals at the BTF \cite{Eigen:2013wka} . 

The high segmentation of the electromagnetic calorimeter in the 
plane transverse to the beam direction assures a spatial resolution of $1/\sqrt12$ = 3 mm which is 
equivalent to an angular resolution at 1.75 m from the target below 2 mrad. 
Energy and angular resolutions obtained with a GEANT4 simulation 
are in agreement with the performance described above. 

\begin{table}[htdp]
\begin{center}
\begin{tabular}{|c|c|c|c|c|c|c|
}
\hline
Density (g/cm$^3$) & $X_0$ (cm) & $R_M$ (cm) & $\tau (ns)$& $\lambda_{peak} (nm)$&L.Y.&dY/dT(\%/C)\\
\hline
7.40 & 1.14 & 2.07 & 40 & 402 &85\%&0.2\\
\hline
\end{tabular}
\end{center}
\caption{Main parameters of the LYSO(Ce) crystals}
\label{tab:LYSO}
\end{table}
 
The average occupancy of the calorimeter varies from 5\% to about 22\% for bunches from $10^4$ to $10^5$ 
positrons. The dependence is not linear due to the splitting of clusters in the inner 
calorimeter region. This is crucial characteristics since it allows efficient
and safe vetoing of events with more than one reconstructed cluster.

\makeatletter{}\section{MC simulation and reconstruction}

To understand the actual sensitivity of the proposed experiment to U bosons, 
a full GEANT4 simulation has been developed.
The simulation describes in detail the segmentation of the calorimeter and 
produces energy deposits for each single crystal.
The magnetic field is considered to be uniform and transverse to the beam 
direction. The spectrometer is modelled as an active volume from 
which the energy of the crossing particles is retrieved without any reconstruction. 
The simulation does not include any passive material and
does not simulate the dumping of the primary beam.

To describe the bunch structure, a simultaneous multi-positron gun was implemented, 
taking into account beam spot size and energy spread in each single burst. 
The simulation uses the GEANT4 low-energy electromagnetic libraries, 
including two photon annihilation, ionization processes, 
Bhabha and Moller scattering, and production of $\delta$-rays.
A custom generator was developed
to simulate the production of the U-boson 
and its eventual decay into $e^+e^-$ or invisible, 
and the three photon annihilation.

The physical properties (lifetime and decay kinematics) of the U boson 
were made dependent on two external parameters, allowing 
the change of the acceptance for different U-boson decay points to be studied. 
A complete mass scan was performed. 

A cluster reconstruction algorithm providing energy and position 
was implemented, 
starting from the energy deposits in each of the calorimeter crystals.
Initially, a seed crystal is identified, 
defined as the one with the maximum energy among all the 
cells in the calorimeter.
A seed is created only if the energy is greater than 
10 MeV to reject clusters 
from low energy radiated photons. 
This condition does not introduce 
inefficiency in reconstructing photons from U boson
production in the mass region under study.  
The cluster is built by summing the energy of all the crystals with 
$E>0.1MeV$ and distance less than 4.6 cm from the seed, corresponding to 2.5 Moliere radii.
All the cells contributing to the cluster are then excluded 
and the algorithm is repeated until no seed is found. 
The cluster position is computed by the energy weighted 
average of the central positions of the cells involved. 
Due to the relatively high number of cells involved 
(up to $\sim$40-50) and the small crystal size a spatial 
resolution as low as 3 mm is achieved. The efficiency for 
cluster reconstruction is $\sim98\%$. 
The present algorithm is not very robust against overlapping 
clusters but is suitable for the purposes of this analysis.

\section{Analysis strategy}

The U boson coupling constant can be determined using the formula 

\begin{equation}
  \frac{\sigma(e^+e^- \to U\gamma)}{\sigma(e^+e^- \to \gamma\gamma)} = 
 \frac{ N( U\gamma)}{ N(\gamma\gamma )} * \frac{Acc(\gamma\gamma)}{Acc(U\gamma)} = \epsilon^2 * \delta , 
\label{eq:eps-calc}
\end{equation}
where $ N( U\gamma) = N( U\gamma)_{obs} - N( U\gamma)_{bkg}$ 
is the number of the signal candidates after the background 
subtraction, $N(\gamma\gamma )$ is the number of observed annihilation events, 
$\delta$ is the $ e^+e^- \to U\gamma$ cross 
section enhancement factor, as described in Figure \ref{fig:ee-ug-sigma-ratio}, 
and $Acc(\gamma\gamma) $ and $Acc(U\gamma)$ are the 
corresponding acceptances for the signal and normalization channels.

\begin{figure}[!htb]

\centering
\includegraphics[width=0.5\textwidth]{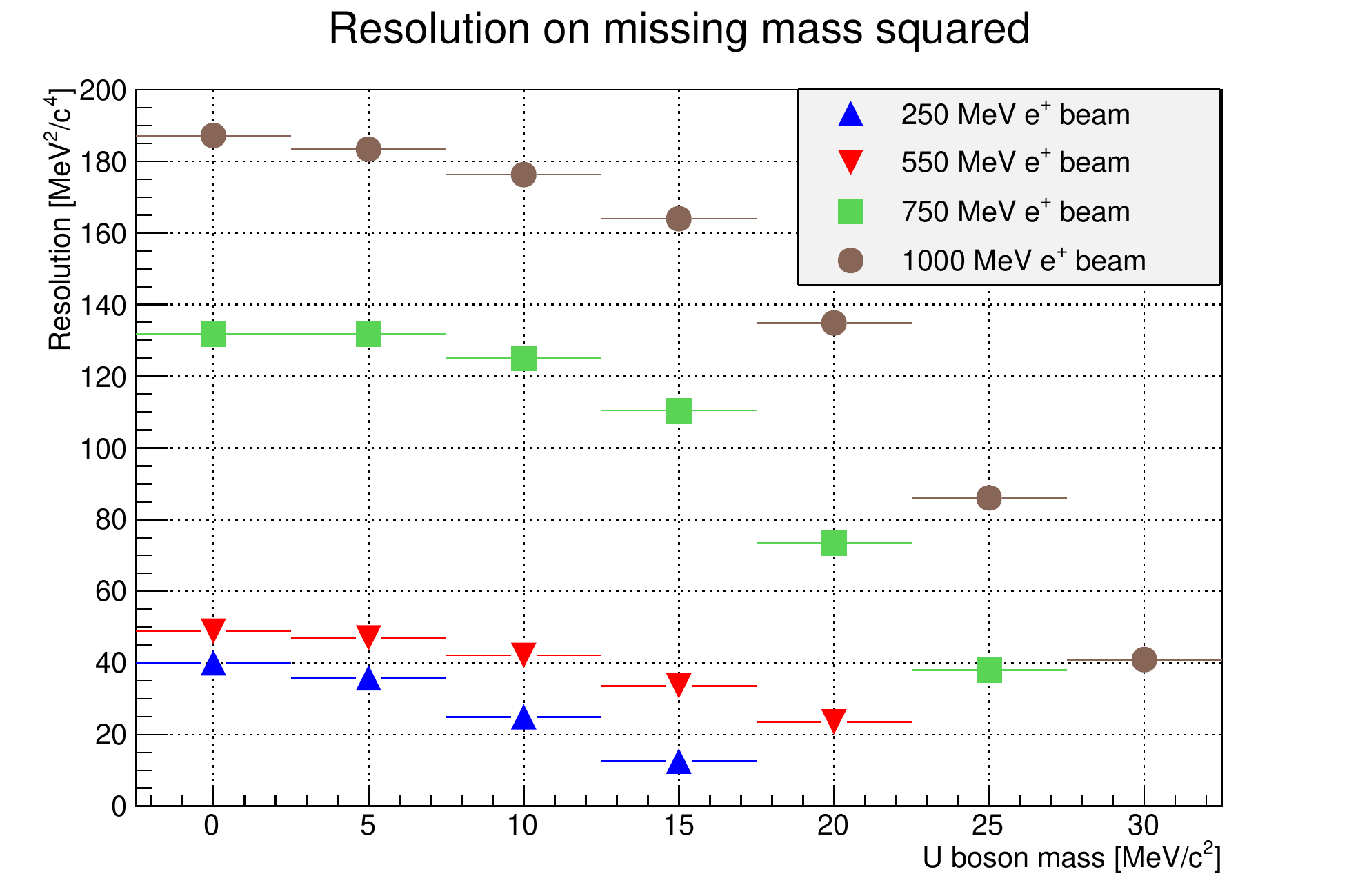}
    \caption{\it Missing mass resolution as a function of the U-boson mass for four different energies of the impinging positron beam}
    \label{fig:mmiss-res-mu}
 \end{figure}

The candidates selection is based on the missing mass squared variable, 
calculated according to formula (\ref{eq:mmiss}), with 
the target electrons assumed to be at rest ($\vec{P}_{e^-} = \vec{0}$) while for the 
beam a nominal momentum of 550 MeV along the Z axis is used. 
The signal region is defined as $ \pm 1.5 \sigma_{M_{miss}^2}$ around the reconstructed value of $M_U^2$,
with resolution $\sigma_{M_{miss}^2}(M_U)$ shown in Figure \ref{fig:mmiss-res-mu}.

A simple and preliminary selection aiming to address the possibility of 
performing a model independent search for a U-boson has been developed. 
It is applied both the events with visible and invisible U-boson decays. 
Since the background estimation does not depend on the U-boson decay channel
the only difference is the change in the acceptance for the two cases. 

The selection cuts applied are the following:

\begin{itemize}
\item Only one cluster in the calorimeter. This cut rejects most of the annihilation events.
\item Cluster energy within $E_{min}(M_U) < E_{Cl}< E_{max}(M_U) $.  
The energy cut is intended to reject low-energy photons from bremsstrahlung 
radiation and to define a maximum energy of the positron in the spectrometer acceptance to avoid pile up.
The energy spectrum of the recoil photons is different for different U masses, 
resulting in $E_{min}(M_U)$ varying over the interval 50-150 MeV 
while $E_{max}(M_U)$ is between 120 and 350 MeV.  
\item  Cluster radius in the calorimeter 5 cm $< R_{Cl}<$ 13 cm. 
This cut reduces the energy leakage and improves $M_{miss}^2$ resolution. 
 \item Positron veto in the spectrometer.

The positron veto cut aims to completely reject the bremsstrahlung gamma 
by detecting the beam positron inside the spectrometer. 
Due to the requirement of $E_\gamma>50$ MeV, the positron is deflected out of the beam by the magnet
and enters the acceptance of the tracker.
The veto condition must not reject the decays of the U bosons, if any. 
In case a positron with energy below 500 MeV is detected, 
$E_{e+}+E_{\gamma}$ has to be lower than 500 MeV. 
In the U boson decay case this cut is equivalent to the soft requirement that the 
electron energy is higher than 100 MeV. However if a high energy electron $E_e > 200 MeV$ is present
the event is kept. The latter is necessary to avoid rejection of events due to pile up.
\end{itemize}

With this selection, an acceptance of $\sim 20\%$ was achieved for U 
boson mass up to 20MeV. 

\subsection{Background estimation}

The U boson sensitivity is limited by the single-photon background. It was estimated by applying 
the signal selection cuts to the $e^+$-on-target MC sample and counting 
the number of events in the signal region.

In bremsstrahlung events, the sum of the energies of the positron track 
and the cluster should be equal to the beam energy since the target is very thin. 
In $\gamma\gamma$ events the final state has only two clusters and the sum of their energies
 is equal to the beam energy.

The bremsstrahlung, the process with highest cross section, 
leads to production of many low energy photons emitted mostly at small angles. 
For carbon target and 550 MeV beam energy, 
about $5\times10^2$ photons with energy 
$>1$ MeV are produced for each annihilation interaction 
(see Figure \ref{fig:cross-section}). 
The calorimeter has a central hole with an aperture of 
1.5 degrees minimizing the sensitivity to these photons. 

\begin{figure}[htb]
  \begin{minipage}[t]{0.48\textwidth}
\centering

     \centering\includegraphics[width=\textwidth]{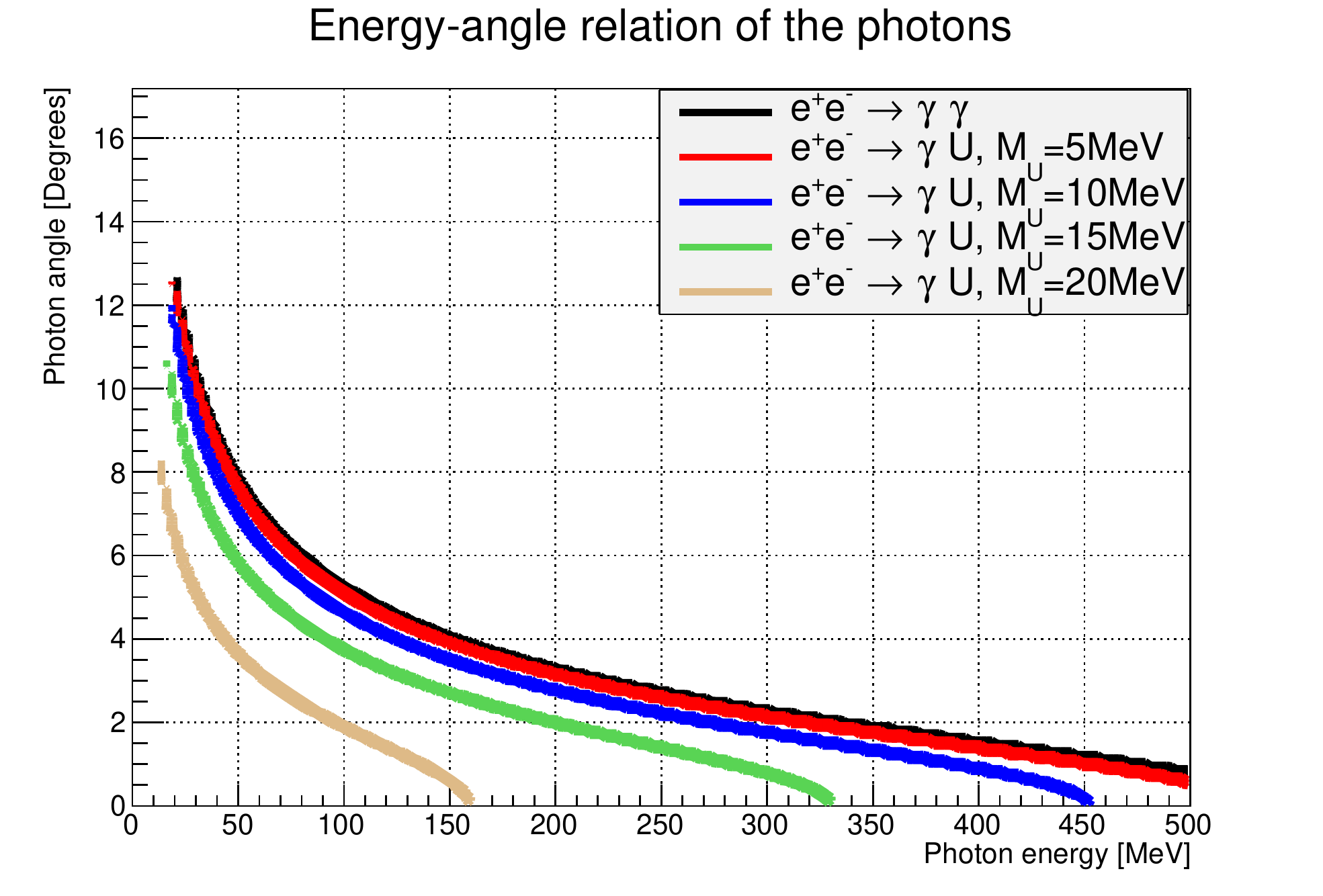}
    \caption{\it Gamma energy versus the opening angle for $e^+e^- \to \gamma\gamma$ or $\to U\gamma$ for different U boson masses. Beam energy is 550 MeV}
    \label{fig:eg-vs-ang}

  \end{minipage}\hfill
  \begin{minipage}[t]{0.48\textwidth}
\centering

\includegraphics[width=\textwidth]{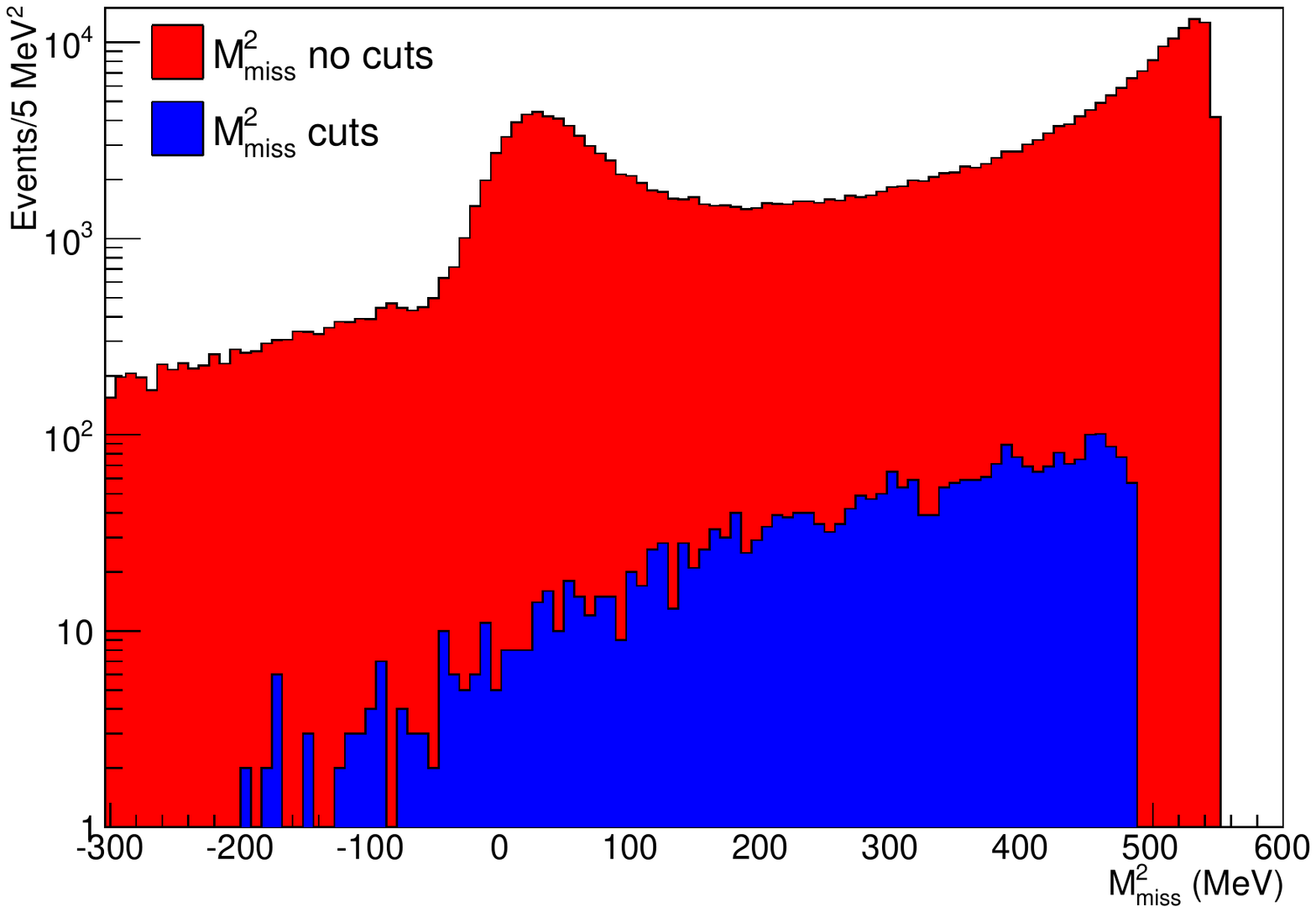}
\caption{\it $M_{miss}^2$ for background events. In red with no selection cuts in blue after all cuts applied.}
\label{fig:MMissBG}

  \end{minipage}
\end{figure}

Because of the closed kinematics the energy and the angle of the annihilation photons
are correlated.  
Due to detector resolution the kinematical region allowed for annihilation 
photons overlaps that for a low mass U boson (see Figure \ref{fig:eg-vs-ang}
).
To reduce background of this type, a veto on extra clusters in the calorimeter 
is applied, leading to a negligible contribution from $\gamma\gamma$ final states. 
The energies of photons produced by synchrotron radiation are very low, 
at most reaching the hard X-ray region ($\sim$10 KeV). 
Nevertheless the deflection of the intense beam can give rise to 
several thousand of these photons, slightly worsening the calorimeter energy resolution. 

An additional background originates from the process $e^+e^- \to \gamma\gamma\gamma$. 
When just a single photon is detected by the calorimeter the reconstructed missing mass is equal to 
the invariant mass of the two missing photons and lies in the signal region. 
That scenario represents an irreducible background when the two photons
fly through the inner hole of the calorimeter. Initial state radiation (ISR) with energy
less than 5 MeV decreases the visible $e^+e^-$ invariant mass from 23.7 MeV to 23.6 MeV and 
is taken into account through simulating the 1\% beam energy spread. The emission of a hard 
photon was studied using CalcHEP. The relative cross section was obtained to be
$\sigma(e^+e^- \to \gamma\gamma\gamma)/\sigma (e^+e^- \to \gamma\gamma)_{E(\gamma)>5MeV} = 7.5*10^7 pb/1.8*10^9 pb = 4.2\%$ 
and the generated events were traced through the experimental setup to obtain the remaining background.

Other background processes like Bhabha scattering and pile up of annihilation events
are included in the background estimation through the GEANT4 simulation of the interactions of the 
primary positron beam. Double annihilation events produce contain extra clusters and due to 
energy/angle relation are additionally suppressed with respect to single ones.

Figure \ref{fig:MMissBG} shows the distribution of the simulated background events
before (red) and after (blue) the described selection.
The annihilation background peaks at $M_{miss}^{2}=0$, the bremsstrahlung 
is located in the region of high $M_{miss}^2$ values while 
the three photon background populates the entire region

\subsection{Positron flux measurement}

The total number of annihilation events ($N(\gamma\gamma )/Acc(\gamma\gamma)$), 
which are used for normalization, can be determined in two independent ways. 
The first is to  
exploit the active target for the measurement of the beam flux and  use 
the known value of $\sigma_{e^+e^-\to\gamma\gamma}$. 
The signal produced 
by the active target is proportional to the number of positrons crossing the target itself.
Its calibration can be performed by shooting the beam directly onto the 
calorimeter through the target, measuring the energy deposit and comparing 
it with the signal from the active target. 

Using this curve the number of primary positrons, and therefore the total flux, 
can be measured in each single bunch.
 Then the number of annihilations can be calculated as 
\begin{equation}
	N_{\gamma\gamma}^{tot} = \frac{N_{\gamma\gamma}}{Acc_{\gamma\gamma}} = Flux(e^+)\cdot \sigma_{\gamma\gamma},
\end{equation}
since the cross section for the annihilation process 
($\sigma_{\gamma\gamma}$) is known with very good precision (Figure \ref{fig:cross-section}). 

\begin{figure}[!htb]
  \begin{minipage}[t]{0.48\textwidth}

\includegraphics[width=\textwidth]{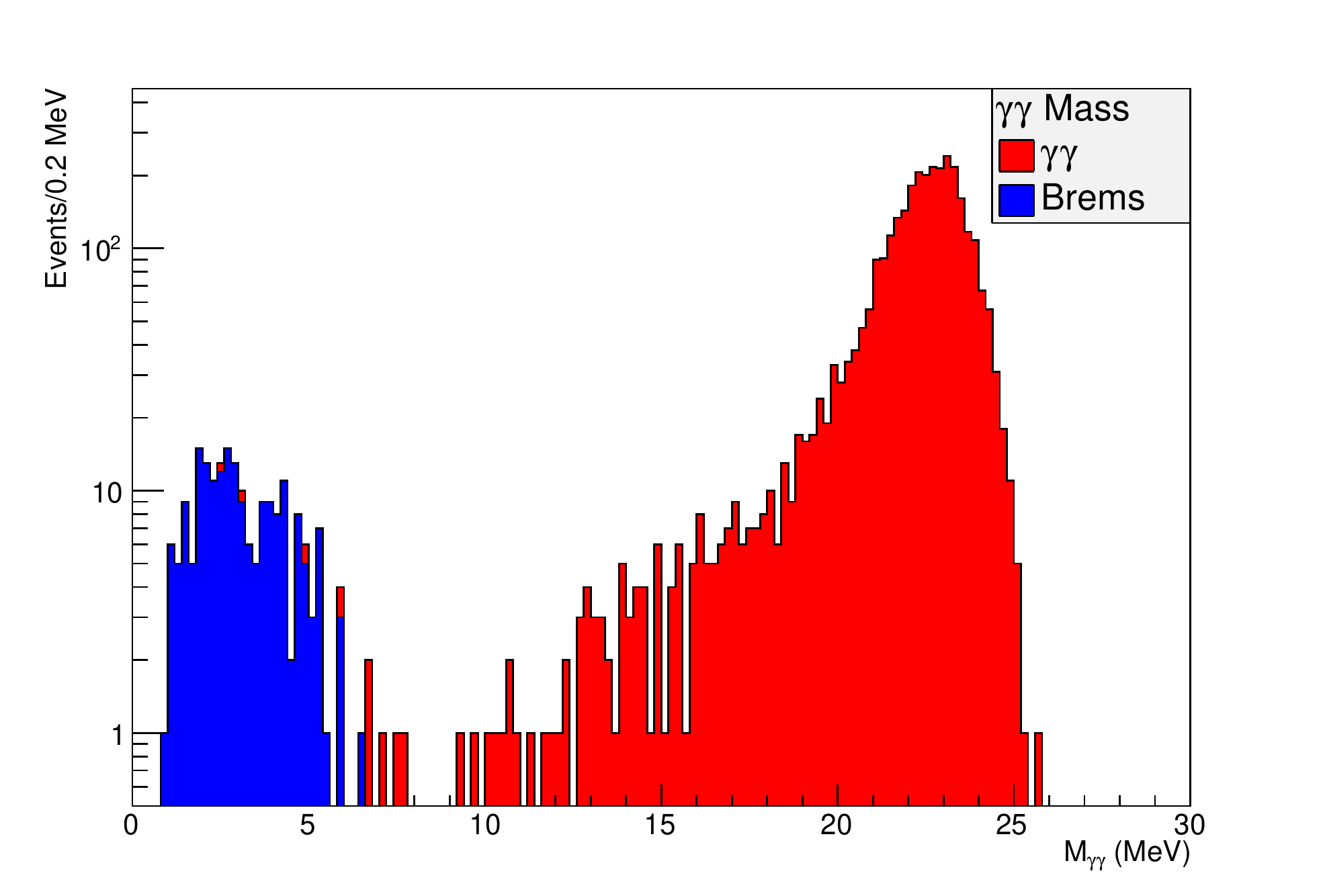}
\caption{\it $M_{\gamma\gamma}$ mass distribution. }
\label{fig:mgg}

  \end{minipage}\hfill
  \begin{minipage}[t]{0.48\textwidth}
  \includegraphics[width=\textwidth]{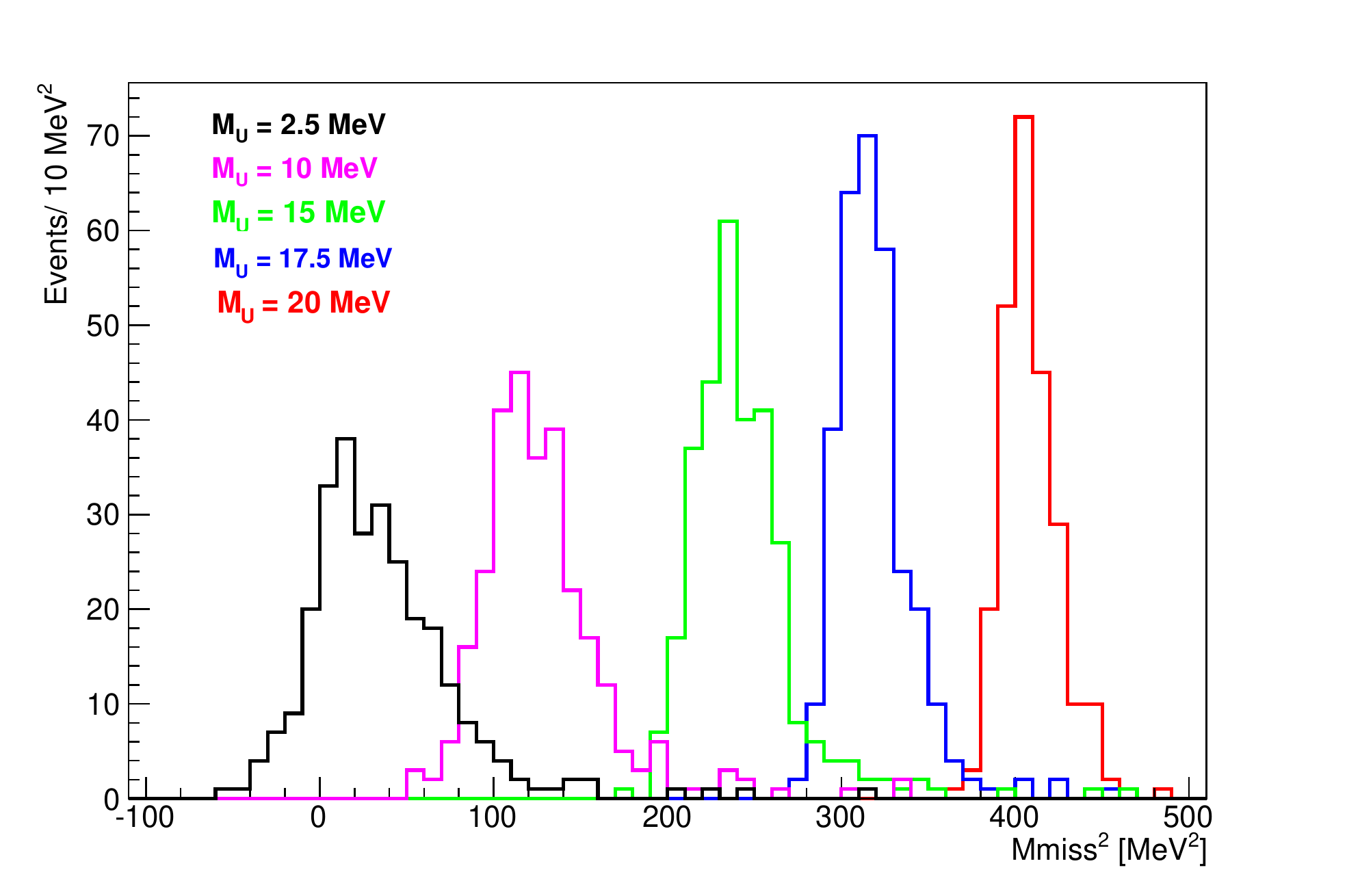}
  \caption{\it Missing mass squared distribution as a function of U-boson mass.}
  \label{fig:Mmiss-U}
  \end{minipage}
\end{figure}

Alternatively, direct reconstruction of the $e^+e^-\to \gamma\gamma$  
annihilation events can be performed.
For the selection of $\gamma\gamma$ events the same geometrical 
cuts for signal events are used. 
Two clusters in the calorimeter are required,  each 
with reconstructed energy 100 MeV $<E_{Cl}<$ 400 MeV and 
radius $5 cm < R_{Cl} < 13 cm $. 
The $\gamma\gamma$ invariant mass is reconstructed assuming 
the particles come from the target Z position and using the formula

\begin{equation}
M_{\gamma\gamma}=\frac{\sqrt{[(X_{\gamma 1}-X_{\gamma 2})+(Y_{\gamma 1}-Y_{\gamma 2})]E_{\gamma 2}E_{\gamma 2}}}{Z_{EMcal}-Z_{Target}}.
\end{equation}

In Figure \ref{fig:mgg} the distribution of $M_{\gamma\gamma}$  from MC is shown. 
The genuine $\gamma\gamma$ events (in red) peak at the centre of mass energy of 
the $e^+e^-$ pair while the negligible bremsstrahlung background (in blue) is 
situated at small $M_{\gamma\gamma}$. 
The expected contamination from bremsstrahlung processes in the signal 
region is well below the 0.1\% level. 
The resolution on $M_{\gamma\gamma}$ is found to be 1 MeV.
Using the GEANT4 simulation and the invariant mass cut 20 MeV $<M_{\gamma\gamma}<$ 26 MeV, 
the acceptance for this selection has been 
measured to be $\sim7\%$, 
with a calorimeter geometrical acceptance for two photons of $\sim17\%$. 
The expected precision in the measurement of the flux is 
dominated by the value of the annihilation cross section, 
since the statistical error on $\gamma\gamma$ sample is $\sim$0.05 \%.

Combining the two results for the number of primary positrons, 
the cross section for the annihilation process ($\sigma_{\gamma\gamma}$) can be measured as a by-product.
This will help to cross check the reliability of the obtained flux, minimizing the systematics. 

\subsection{Sensitivity}
\begin{figure}[htb]
  \begin{minipage}[t]{0.47\textwidth}
    \centering\includegraphics[width=\textwidth]{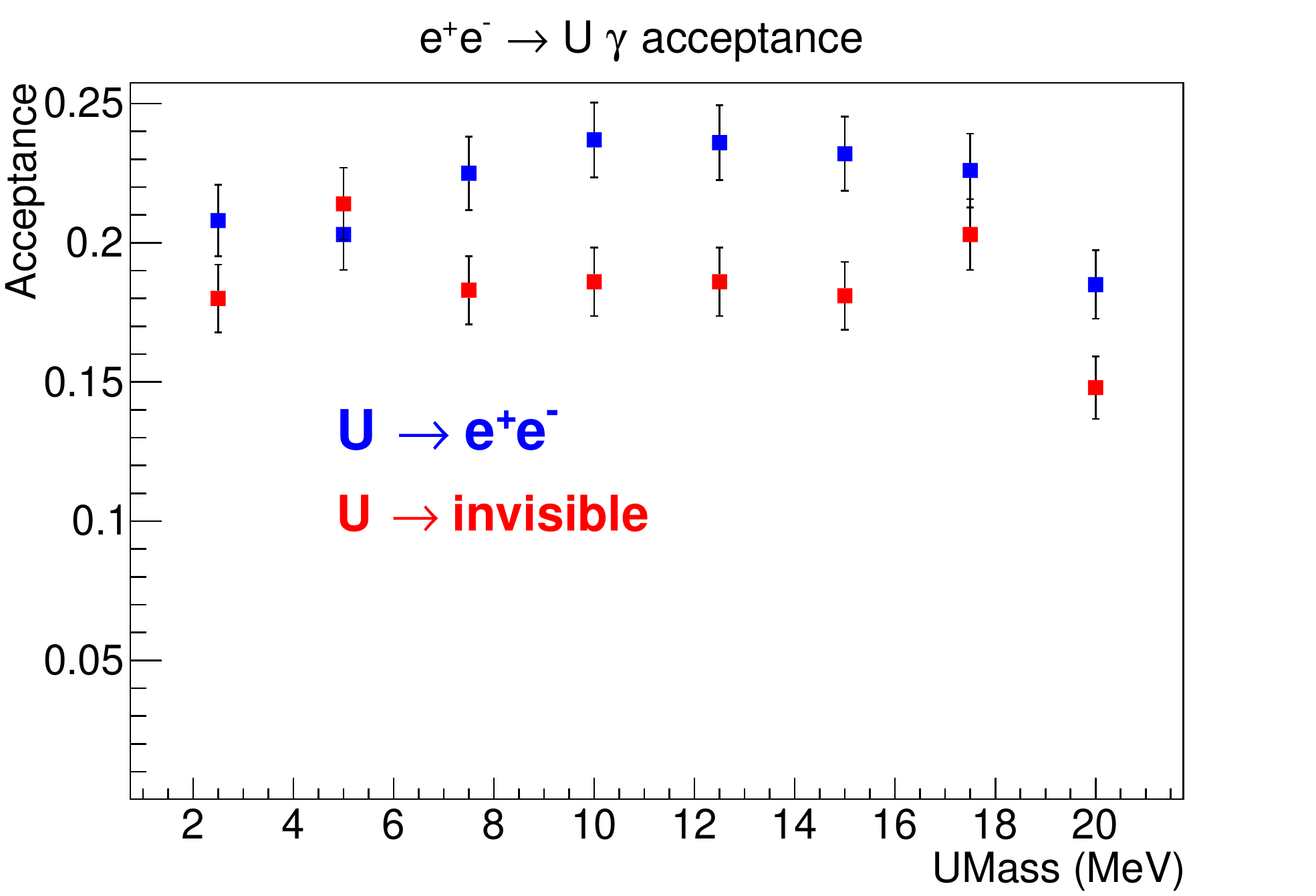}
    \caption{\it Acceptance for U boson detection as a function of its mass}
    \label{fig:u-acc}
  \end{minipage}\hfill
  \begin{minipage}[t]{0.50\textwidth}
    \centering\includegraphics[width=\textwidth]{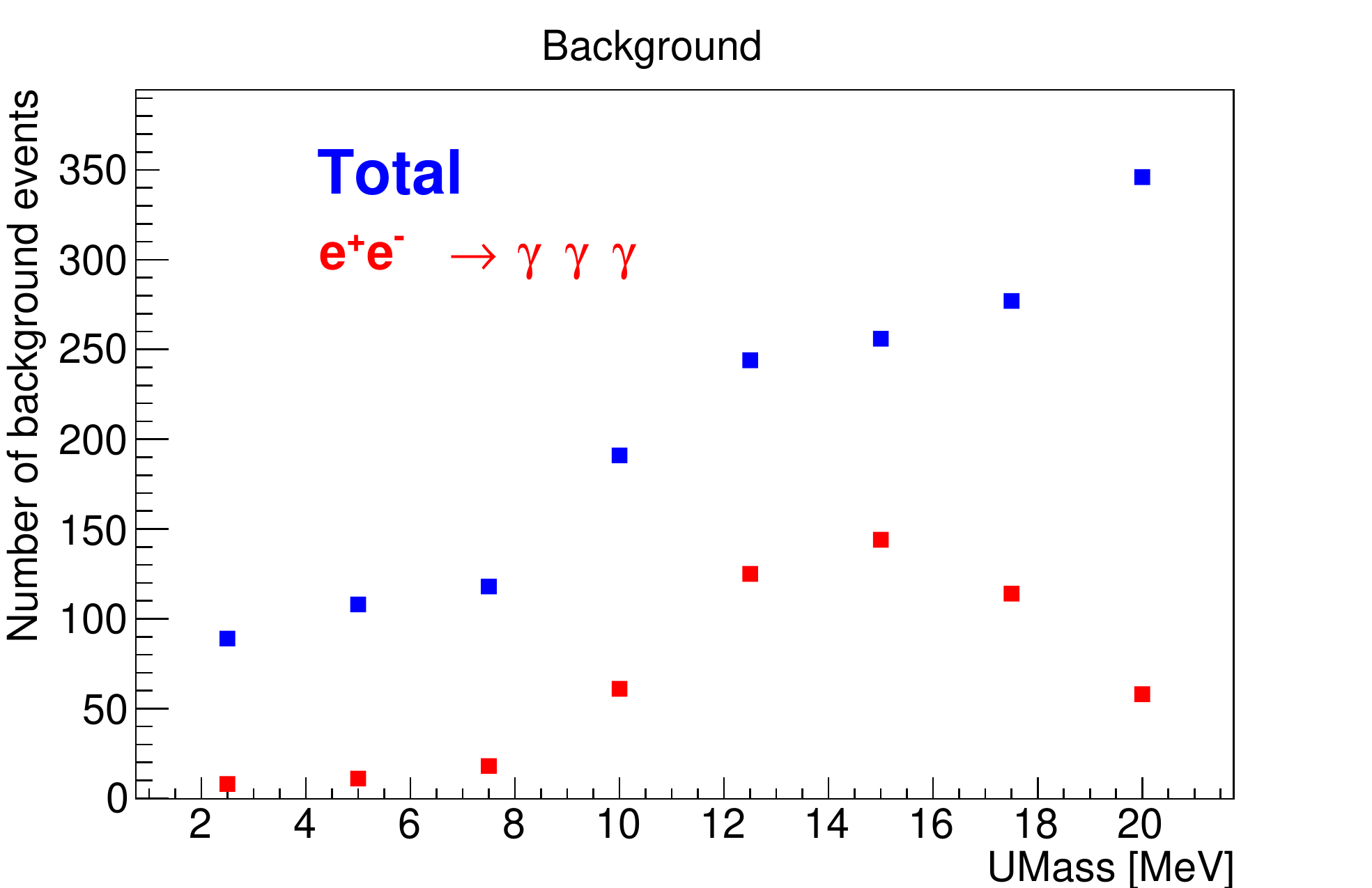}
    \caption{\it Background contribution as a function of the U boson mass }
    \label{fig:u-bkg}
  \end{minipage}
\end{figure}

With the described experimental setup and simulation technique
$10^{11}$ positrons on target were generated ($10^7$ events each with $10^4$ positron 
and $10^8$ events each with $10^3$ positron) in order to study the effect of pile up events. 
In addition, samples of 1000 events were generated for U-boson masses 
2.5, 5, 7.5, 10, 12.5, 15, 17.5, 20 MeV, with a single U boson 
 per bunch with $2*10^3$ positrons, corresponding to a $4*10^4$ positrons per 40 ns bunch. 
This assumes that the detector is able to match clusters and tracks within 2 ns.  
The missing mass distributions for some of the masses are shown in Figure \ref{fig:Mmiss-U}.
The acceptance (Figure \ref{fig:u-acc}) and the expected number of background events (Figure \ref{fig:u-bkg}) 
were obtained by running the events through the selection.  

\begin{figure}[t]
  \begin{minipage}[t]{0.54\textwidth}
  \centering \includegraphics[width=\textwidth]{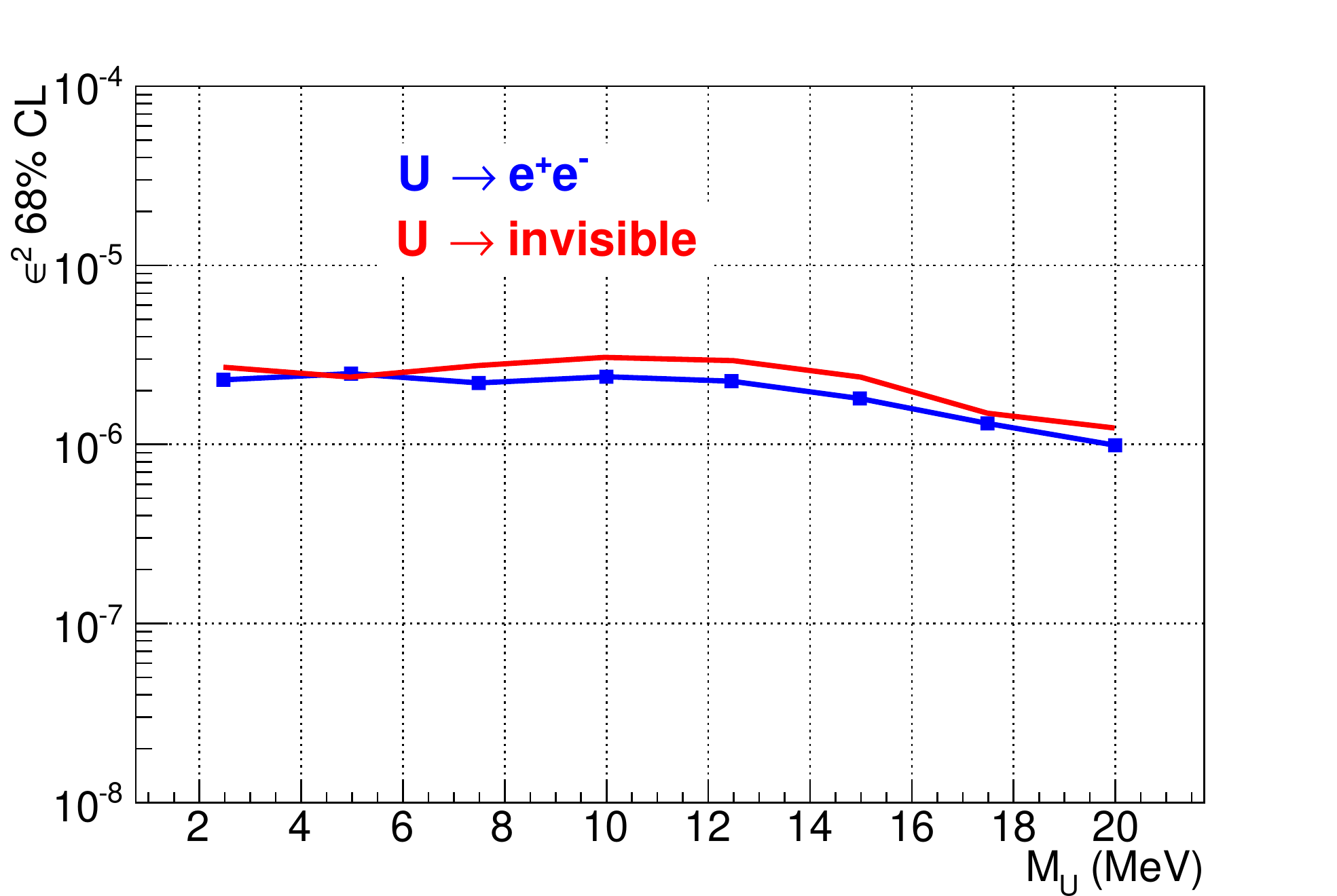}
  \caption{Expected exclusion limits in the $\epsilon - M_U$ plane in case of no signal}
\label{fig:u-excl}
  \end{minipage}\hfill
  \begin{minipage}[t]{0.42\textwidth}
    \centering\includegraphics[width=\textwidth]{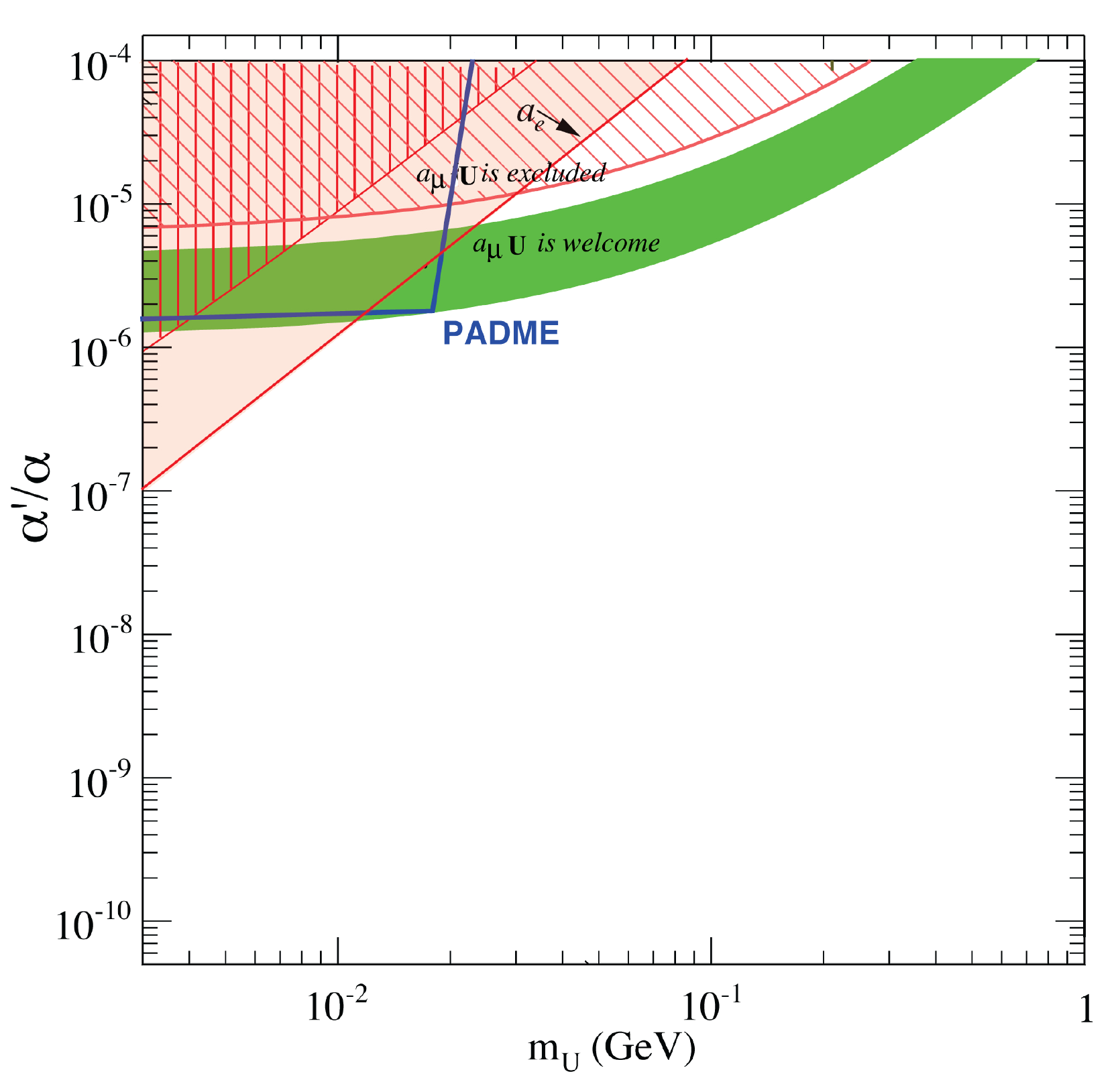}
    \caption{\it Expected exclusions in the invisible channel compared with 
  the band of values preferred by current $g_{\mu}-2$ discrepancies. }
    \label{fig:u-excl-inv}
  \end{minipage}
\end{figure}

The background was further scaled by a constant factor of 400 
to account for one year of running of the experiment with 60\% efficiency,  
$4*10^4$ positrons per bunch, corresponding to a total of $4*10^{13}$ positrons on target. 
Under the assumption of no signal, an upper limit on the coupling $\epsilon$ can be set, 
using the statistical uncertainty on the background. The expected exclusion region is shown 
in Figure \ref{fig:u-excl} for both the visible and the invisible channels.

\makeatletter{}\section{Conclusions and discussion}
Searches for dark photons are well motivated by recently observed phenomena.
The analysis presented in this paper shows that such a search can be performed exploiting  the present 
linac of the DA$\Phi$NE facility in just one year of running. 
The expected sensitivity $\epsilon^2 \sim 10^{-6}$ is common both to the 
visible and invisible U boson decays and 
lies exactly in the region, preferred by the 
$g_{\mu} -2 $ data, as seen in Figure \ref{fig:u-excl-inv}. 
It partially overlaps with the result of the combination of
the measurements of $(g_e - 2)$ and $\alpha_{EM}$, 
exploiting tenth-order calculations of the QED contributions.
As discussed in section (\ref{sec:pres-stat}),
that region is still of high interest since no direct 
search for a U-boson has been performed accessing it and because of 
the recent shift by more than 7$\sigma$ \cite{bib:g-2-error} 
of the theoretical prediction for $g_e$ due to an error in the calculations. 
The proposed experiment could be the first to directly constrain the invisible channel. 
An upgrade of the BTF positron beam energy to 750 MeV will extend the 
sensitivity to higher U-boson mass. 

At present there are a few experiments in different phases of preparations devoted to the search 
for dark photons - 
VEPP3 \cite{bib:vepp3}, DarkLight \cite{bib:darklight},  HPS \cite{bib:HPS}, and BDX\cite{bib:BDX}. 
Among them, only VEPP3 is sensitive to invisible decays in 
the same region of parameter space. 
While the general approach is similar, there are few crucial differences.
The VEPP3 proposal aims to use a storage ring which limits the beam energy to a fixed 
value of 500 MeV, while the DA$\Phi$NE linac could be able to 
provide positrons with variable beam energy from 250 MeV up to 750 MeV 
and a beam spread of 1\%, 
thus extending the accessible region of U-boson masses to $\sim$ 27.5 MeV. 
Moreover, the presence of an 
active target allows the determination of the interaction vertex at the mm level, 
leading to a reduction of the background through improvement of the missing mass resolution.
Another advantage of the experiment proposed in this paper is the relatively small calorimeter,
since it is placed at a distance from the target about three times smaller with respect to the 
VEPP3 one, which minimizes the cost and timescale of the experiment. 
The presence of a magnetic spectrometer, 
apart from further reducing the background by measuring 
the momentum of the positrons that had emitted hard bremsstrahlung photons,
could possibly allow the 
direct study of visible $U \to e^+e^-$ decays. 
This channel makes the experiment 
sensitive to the
bremsstrahlung production of U bosons, 
thus pushing the accessible parameters region to higher masses. 
While providing an interesting benchmark possibility with the comparison of the results from
visible and invisible searches done simultaneously, this falls outside 
the scope of the present paper and will be a subject of a future study. 

\section*{Acknowledgments}
The authors would like to thank Antonella Antonelli, Fabio Bossi, 
Paolo Valente, Matthew Moulson, 
Tommaso Spadaro, Sarah Andreas and the LNF NA62 large angle veto group  
for encouraging those activities and for all the useful discussions. 
We would like also to thank the linac team at LNF for the 
information on present status and possible upgrade of BTF. 
The results were obtained exploiting the LNF computing facilities.

\makeatletter{}


\begin{thebibliography}{99}

\bibitem{bib:kin-mixing}
  B. Holdom, Phys. Lett. B {\bf166}, 196 (1986) \\
  P. Galison and A. Manohar, Phys. Lett. B {\bf136}, 279 (1984).


\bibitem{bib:pamela2008}
O. Adriani  {\it et al.} [PAMELA Collaboration], 
Nature {\bf 458}, 607 (2009). 

\bibitem{bib:fermi-pos}
M. Ackermann  {\it et al.} [Fermi LAT Collaboration],
Phys. Rev. Lett. {\bf 108}, 011103 (2012).

\bibitem{bib:ams-pos}
M. Aguilar  {\it et al.} [AMS Collaboration], 
Phys. Rev. Lett. {\bf 110}, 141102 (2013).

\bibitem{bib:pamela-antip}
 O. Adriani  {\it et al.} [PAMELA Collaboration],
 Phys. Rev. Lett. {\bf 105}, 121101 (2010).

\bibitem{bib:g-2-discrepancy}
J. Beringer {\it et al.} [Particle Data Group], 
Phys. Rev. D{\bf 86}, 010001 (2012).


\bibitem{bib:u1-gauge}
P.~Fayet, 
Phys.\ Lett.\ B\ {\bf 675}, 267 (2009).


\bibitem{bib:calchep}
   A.Pukhov  {\it et al.}, Preprint INP MSU 98-41/542,arXiv:hep-ph/9908288 \\
   A.Pukhov, e-Print Archive: hep-ph/0412191.
 
\bibitem{bib:sarah-dark} 
S.~Andreas, C.~Niebuhr and A.~Ringwald,
    Phys.\ Rev.\ D {\bf 86}, 095019 (2012).

\bibitem{DUMP1} 
  E.~M.~Riordan
{\it et al.},
    Phys.\ Rev.\ Lett.\  {\bf 59}, 755 (1987).
    
\bibitem{bib:E137-dark}
J. D. Bjorken {\it et al.},
Phys. Rev. D{\bf38}, 3375 (1988). 


\bibitem{E774} 
  A.~Bross, M.~Crisler, S.~H.~Pordes, J.~Volk, S.~Errede and J.~Wrbanek,
    Phys.\ Rev.\ Lett.\  {\bf 67}, 2942 (1991).
    
\bibitem{bib:gsi-pos-peak}
  J. Schweppe {\it et al.}, 
Phys.\ Rev.\ Lett. {\bf 51}, 2261 (1983). 



\bibitem{oai:arXiv.org:1101.4091} 
  H.~Merkel {\it et al.}  [A1 Collaboration],
    Phys.\ Rev.\ Lett.\  {\bf 106}, 251802 (2011).
        
\bibitem{APEX} 
  S.~Abrahamyan {\it et al.}  [APEX Collaboration],
    Phys.\ Rev.\ Lett.\  {\bf 107}, 191804 (2011).
      
\bibitem{bib:bossi-dp}
F.~Bossi, 
Advances in High Energy Physics {\bf 2014}, 891820 (2014).

\bibitem{bib:essig-col}
R.~Essig {\it et al.}, 
JHEP {\bf 1311}, 167 (2013).


\bibitem{Adlarson:2013eza}  
  P.~Adlarson {\it et al.}  [WASA-at-COSY Collaboration],
    Phys.\ Lett.\ B {\bf 726}, 187 (2013).
    
\bibitem{Archilli:2011zc}    
  F.~Archilli, D.~Babusci, D.~Badoni, I.~Balwierz, G.~Bencivenni, C.~Bini, C.~Bloise and V.~Bocci {\it et al.},
    Phys.\ Lett.\ B {\bf 706}, 251 (2012).
      
\bibitem{bib:kloe-phi-u}
D.~Babusci {\it et al.}, 
  Phys.\ Lett.\ B {\bf 720}, 111 (2012).


 \bibitem{oai:arXiv.org:0905.4539} 
  B.~Aubert {\it et al.}  [BaBar Collaboration],
    Phys.\ Rev.\ Lett.\  {\bf 103}, 081803 (2009).
       

\bibitem{bib:ub-g2em}
M. Pospelov,
Phys.\ Rev.\ D {\bf80}, 095002 (2009).



\bibitem{bib:g-2-elec}
D. Hanneke, S. Fogwell Hoogerheide, and G. Gabrielse,
Phys. Rev. A {\bf 83}, 052122 (2011).


\bibitem{bib:alpha-rb}
R.~Bouchendira {\it et al.}, 
Phys.\ Rev.\ Lett.\ {\bf 106}, 080801 (2011).

\bibitem{bib:mQ}
M. Diamond and P. Schuster, 
Phys.\ Rev.\ Lett.\ {\bf 111}, 221803 (2013).


\bibitem{bib:ge-2-qed}
T.~Aoyama {\it et al.}, 
Phys. Rev. Lett. {\bf 109}, 111807 (2012).

\bibitem{bib:dark-review}
R.~Essig {\it et al.}, 
arXiv:1311.0029 [hep-ph]


\bibitem{bib:na62}
F. Hahn {\it et al.} [NA62 Collaboration], http://cds.cern.ch/record/1404985.


\bibitem{bib:toro-new-dm-search}
E. Izaguirre {\it et al.}, 
Phys. Rev. {\bf D} 88, 114015 (2013).


\bibitem{Ghigo:2003gy}
  A.~Ghigo, G.~Mazzitelli, F.~Sannibale, P.~Valente and G.~Vignola,
    Nucl.\ Instrum.\ Meth.\ A {\bf 515}, 524 (2003).

\bibitem{bib:btf-upg}
P.~Valente, ``Possible upgrades of the DAFNE Beam Test Facility'', unpublished

\bibitem{bib:btf-gun-upg}
B.~Buonomo et al., DAFNE note in preparation, (2014).

\bibitem{bib:annih}
   W. Heitler.,  The Quantum Theory of Radiation, Clarendon Press, Oxford (1954).

\bibitem{Eigen:2013wka} 
  G.~Eigen {\it et al.},
    Nucl.\ Instrum.\ Meth.\ A {\bf 718}, 107 (2013).
  
\bibitem{bib:g-2-error}
  T. Aoyama, M. Hayakawa, T. Kinoshita, and M. Nio, 
  Phys. Rev. Lett. {\bf 99}, 110406 (2007).


\bibitem{bib:vepp3}
B. Wojtsekhowski, AIP Conf. Proc. {\bf 1160}, 149 (2009), arXiv:0906.5265 [hep-ex]\\
B. Wojtsekhowski, D. Nikolenko and I. Rachek, arXiv:1207.5089 [hep-ex].

\bibitem{bib:darklight}
M. Freytsis, G. Ovanesyan and J. Thaler, 
JHEP {\bf 1001}, 111 (2010); [arXiv:0909.2862 [hep-ph]].

\bibitem{bib:HPS}
The Heavy Photon Search Collaboration (HPS), 

https://confluence.slac.stanford.edu/display/hpsg/

\bibitem{bib:BDX}
M.~Battaglieri {\it et al.}  [ BDX Collaboration],
  arXiv:1406.3028 [physics.ins-det].

\end{thebibliography}
\end{document}